# Title: Electrically controllable valence-conduction band reversals in helical trilayer graphene

**Authors:** Matan Bocarsly[1†], Indranil Roy[1†], Weifeng Zhi[1†], Li-Qiao Xia[2†], Aviram Uri[2], Yves H. Kwan[3], Aaron Sharpe[4,5], Matan Uzan[1], Yuri Myasoedov[1], Kenji Watanabe[6], Takashi Taniguchi[7], Trithep Devakul[4], Pablo Jarillo-Herrero[2]* and Eli Zeldov[1]*

**Affiliations:**

[1]Department of Condensed Matter Physics, Weizmann Institute of Science, Rehovot 7610001, Israel
[2]Department of Physics, Massachusetts Institute of Technology, Cambridge, MA 02139, USA
[3]Department of Physics, University of Texas at Dallas, Richardson, TX 75080, USA
[4]Department of Physics, Stanford University, Stanford, CA, USA 94305
[5]Stanford Institute for Materials and Energy Sciences, SLAC National Accelerator Laboratory, Menlo Park, CA 94025
[6]Research Center for Electronic and Optical Materials, National Institute for Materials Science; 1-1 Namiki, Tsukuba 305-0044, Japan
[7]Research Center for Materials Nanoarchitectonics, National Institute for Materials Science; 1-1 Namiki, Tsukuba 305-0044, Japan
[†]These authors contributed equally to this work
*e-mail: pjarillo@mit.edu, eli.zeldov@weizmann.ac.il

**Abstract:** In moiré graphene systems, electronic interactions lift spin and valley degeneracies, leading to symmetry-broken ground states. In helical trilayer graphene (HTG), we uncover a distinct interaction-driven mechanism in which the roles of sublattice-polarized valence and conduction bands are cyclically reversed. Using scanning nano-SQUID magnetometry, we detect a series of sharp magnetic signatures consistent with seesaw-like transitions, where occupied and unoccupied valence and conduction bands interchange repeatedly with doping, accompanied by a novel form of magnetic hysteresis. These transitions occur entirely within metallic regimes and leave only weak fingerprints in transport measurements. Self-consistent Hartree-Fock calculations reveal that interactions reorganize all eight low-energy flat bands, driving abrupt changes in orbital magnetization. Our results establish HTG as the first system where electronic interactions provide doping-controlled access to all three internal degrees of freedom—spin, valley, and sublattice—introducing a new class of correlated phase transitions.

**Main Text:**

Graphene-based moiré systems provide a rich platform for exploring correlated quantum phases, enabled by the emergence of topological flat bands and strong electron-electron interactions. In these systems, each electron is characterized by three intertwined quantum degrees of freedom—spin, valley, and sublattice polarization—which together define eight low-energy flat bands near charge neutrality. Interaction-driven symmetry breaking has been extensively studied in the spin and valley sectors, resulting in correlated phases such as spin-polarized [1–3], valley-polarized [4–11], and inter-valley coherent states [12,13]. These phenomena typically occur within either the conduction or valence bands alone. In contrast, symmetry breaking states involving both valence and conduction bands of a single spin-valley sector have yet to be observed.

Here, we demonstrate that in helical trilayer graphene (HTG), electronic interactions can reorganize the entire low-energy band manifold in response to doping. Rather than operating within a fixed valence-conduction band hierarchy, the system undergoes a sequence of reconfigurations in which the roles of filled and empty sublattice-polarized bands are repeatedly reversed. This reveals a new interaction-driven mechanism—one that engages all three electronic degrees of freedom in a unified and tunable way. Our findings establish HTG as a unique platform for realizing emergent quantum phases shaped by the full spin-valley-sublattice coupling landscape.

HTG consists of three graphene monolayers twisted sequentially by the same angle (Fig. 1B inset), forming a nontrivial electronic structure with flat bands of distinct topological character [14–22]. Lattice relaxation in HTG leads to moiré-periodic domains that inherently break $xy$-inversion symmetry, producing a super moiré structure composed of ordered h-HTG and $\bar{\text{h}}$-HTG domains that are mutual $xy$-inversion counterparts [14]. A recent electrical transport study revealed that as the twist angle approaches the magic angle, $\theta_\text{m} \cong 1.8°$, HTG exhibits an anomalous Hall effect (AHE) at odd-integer moiré fillings, indicating the presence of topological bands and correlation-driven spontaneous time-reversal symmetry breaking [15]. While Chern insulators have been identified among the incompressible states in various moiré systems [4–6,9–11,23–28], the nature of correlated electronic phases in the compressible regime is less well-characterized. This is largely because few experimental techniques are sensitive to phase transitions in metallic states.

In systems lacking $xy$-inversion symmetry, moiré bands acquire nonzero Berry curvature and exhibit orbital magnetization. This magnetization has two components (Supplementary Materials) [11,29,30]: the self-rotation magnetization, arising from the internal angular momentum of electronic states about their center of mass; and the Chern magnetization, which is sensitive to the position of the chemical potential and reflects the collective motion of electrons in topological bands. Since both components are present even in gapless phases, orbital magnetometry offers a powerful means to detect electronic phase transitions inaccessible by conventional transport.

Here we report the first magnetic imaging study of HTG, revealing a sequence of sharp magnetic transitions in the compressible regime upon electron doping, with a nontrivial dependence on carrier density and transverse displacement field. Whereas interaction-driven transitions in electron-doped moiré systems are conventionally understood as occurring within a set of degenerate conduction bands, we show that this framework breaks down in HTG and cannot account for the observed experimental signatures. Self-consistent Hartree-Fock (scHF) calculations identify these transitions as interaction-driven reorganizations of the entire low-energy band manifold, involving all eight flat bands, mediated by controlled flipping of sublattice polarization. As doping increases, the system undergoes repeated reversals in the roles of filled and empty sublattice-polarized Chern bands, minimizing the combined kinetic and Coulomb energies. Although these transitions occur in metallic states, they are accompanied by abrupt jumps in Chern magnetization, revealing a mechanism previously unobserved in moiré systems. Moreover, these sublattice transitions—effectively interchanging valence and conduction bands—give rise to an exotic form of hysteresis, fundamentally distinct from previously observed hysteretic Chern band systems [4–6,9–11,27,28,31]. These results position HTG as a versatile platform for exploring unconventional interaction-driven reordering of electronic bands and emergent topological phenomena.

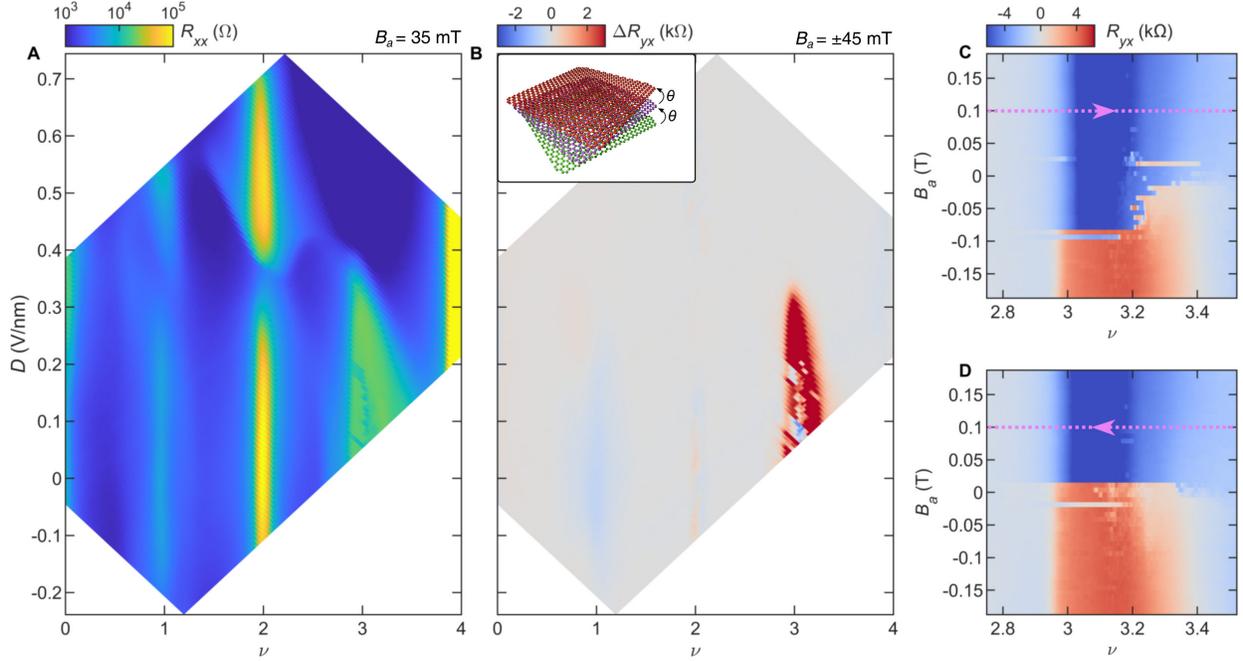

**Fig. 1. Transport measurements in HTG. A,** $R_{xx}$ versus $\nu$ and $D$ measured at $B_a = 35$ mT and $T = 300$ mK in device A. Resistance peaks appear at charge neutrality and at the moiré band gap ($\nu = 4$), which are consistent with single particle physics, while interaction-induced peaks appear at $\nu = 1, 2, 3$. **B,** $\Delta R_{yx}$ measured at $B_a = \pm 45$ mT as a function of $\nu$ and $D$, revealing AHE near $\nu = 1, 3$. The weak signal visible at $\nu = 2$ results from finite mixing between $R_{xx}$ and $R_{yx}$. Inset: schematic of helically twisted trilayer graphene. **C-D,** $R_{yx}$ in the vicinity of $\nu = 3$ at low $B_a$ displaying pronounced hysteresis upon sweeping $\nu$ up (**C**) and down (**D**).

**Transport measurements**

Dual-gated magic-angle HTG devices encapsulated in hexagonal boron nitride (hBN) were fabricated using the cut-and-stack technique (Supplementary Materials). The *dc* voltages $V_{tg}^{dc}$ and $V_{bg}^{dc}$ applied to the top and bottom gates allow for independent control of the carrier density $n$ and displacement field $D$ (Supplementary Materials), while measuring four-terminal transport properties. Figure 1A shows the longitudinal resistance $R_{xx}$ of device A as a function of filling factor $\nu = 4n/n_s$ and $D$ at a temperature $T = 300$ mK and applied magnetic field $B_a = 35$ mT (where $n_s$ corresponds to the density of four electrons per moiré unit cell). As reported previously [15], resistive states are discerned at $\nu = 1, 2, 3$ for a wide range of $D$, generally associated with the formation of flavor-symmetry-broken states within the four-fold conduction bands. At $D \cong 0.35$ V/nm, the resistive states disappear and reappear at larger $D$ for $\nu = 1, 2$. Additionally, AHE is observed, as seen in Fig. 1B which shows the difference $\Delta R_{yx}$ taken between $R_{yx}$ at $B_a = \pm 60$ mT. AHE hotspots appear near $\nu = 1$ and 3 but apparently not at $\nu = 2$. The strongest AHE appears near $\nu = 3$ but without full quantization, similar to a previous report [15]. To explore this state further, we measure $R_{yx}$ upon sweeping $\nu$ up (Fig. 1C) and down (Fig. 1D) at different small $B_a$. The hysteresis pattern upon sweeping $\nu$ is reminiscent of similar behavior observed at odd integer fillings in magic angle twisted bilayer graphene (MATBG) [4–6,10,11,28].

**Magnetic imaging and phase diagram**

To investigate local magnetism, we utilize a scanning superconducting quantum interference device fabricated on the apex of a sharp pipette (SQUID-on-tip, SOT). An Indium SOT [32] of diameter $d \cong 180$ nm and a field sensitivity down to 10 nT/Hz$^{1/2}$, is scanned at a height of $h \cong 180$ nm above the sample surface (Fig. 2A) at $T = 300$ mK in the presence of small applied $B_a$ (Supplementary Materials). A small *ac* voltage $V_{tg}^{ac}$ of 50 mV rms applied to the top gate modulates the carrier density by a small $n^{ac}$ and the resulting *ac* stray magnetic field $B_z^{ac} \sim n^{ac}(dB_z/dn)$ is imaged, which reflects the corresponding local differential *ac* magnetization (Supplementary Materials).

Figure 2B shows a map of $B_z^{ac}$ as a function of $\nu$ and $D$ at a fixed SOT position in device B. This device exhibits no signatures of h-HTG and $\bar{\text{h}}$-HTG domains [14,20] (see Fig. S1 for discussion of domain structure in three investigated samples). The $\nu$-$D$ magnetic phase diagram reveals a rich and complex structure, characterized by four prominent lines of $B_z^{ac}$ peaks with nontrivial shapes, labeled I-IV (see Figs. S2 and S3 for additional locations, samples, and magnetic fields). Additional weaker features are resolved at integer fillings. At low $D$, the peaks appear in the compressible regions between $\nu = N$ and $N + 1$, for $N = 0,1,2,3$. As $D$ is increased, the peak trajectories move towards lower densities, eventually crossing through the integer filling factors ($\nu = 1, 2$ and 3). A detailed analysis reveals that these $B_z^{ac}$ peaks correspond to sharp, large changes in local magnetization $M_z$, with magnetization steps reaching $\Delta M_z \cong 3\ \mu_B$ per moiré unit cell (Supplementary Materials, Fig. S4). Observation of local magnetization establishes that time reversal symmetry is spontaneously broken. Furthermore, the large magnetization steps—far exceeding what can be attributed to spin magnetization—identify the $B_z^{ac}$ peaks as orbital in origin, reflecting interaction-driven electronic reordering that occurs within the metallic state. Analysis of the excitation-amplitude dependence further shows that the phase transitions are second-order (Fig. S5).

Notably, the most pronounced features of $R_{xx}$ peaks at integer fillings (Fig. 1A) have only weak signatures in the magnetization data (Fig. 2B), and conversely, the prominent peak lines in the local magnetic response lack clear counterparts in transport. However, closer inspection resolves weak features and minima in $R_{xx}$ that trace trajectories similar to those of the magnetic peaks (Fig. S2F). The magnetic behavior observed in Fig. 2B stands in stark contrast to that of MATBG, the prototypical moiré topological flat-band system. In MATBG, the strongest differential orbital magnetization at zero $B_a$ occurs in the Chern gaps [10,11], whereas here, it emerges in compressible regions. The weak fingerprints of these electronic phase transitions in transport underscore the unique sensitivity of local SOT measurements in detecting symmetry-breaking transitions in metallic states, as discussed next.

**Hartree-Fock seesaw sublattice switching**

Since the magnetic peak trajectories I-IV are distinct and reproducible across different samples (Fig. S2) and magnetic fields (Fig. S3), they provide a robust basis for developing a phenomenological understanding of the interaction-induced phase diagram of HTG. The large $R_{xx}$ at charge naturality is consistent with a gapped state in which the fourfold-degenerate valence bands are fully occupied, and the fourfold-degenerate conduction bands are empty. A natural null hypothesis is that the magnetic peaks arise from interaction-driven transitions occurring solely within the conduction bands, with the valence bands remaining inert, as commonly observed in other moiré systems such as MATBG. However, this picture fails to explain the experimental data in HTG in Fig. 2B. Our quantitative analysis (Fig. S9 and Supplementary Materials) shows that this null hypothesis can reproduce only two compressible magnetization jumps, in contrast to the four robust peaks observed experimentally. We therefore conclude that conventional conduction-band-only transitions cannot account for the observed phenomenology, pointing to a qualitatively different interaction-driven mechanism in HTG.

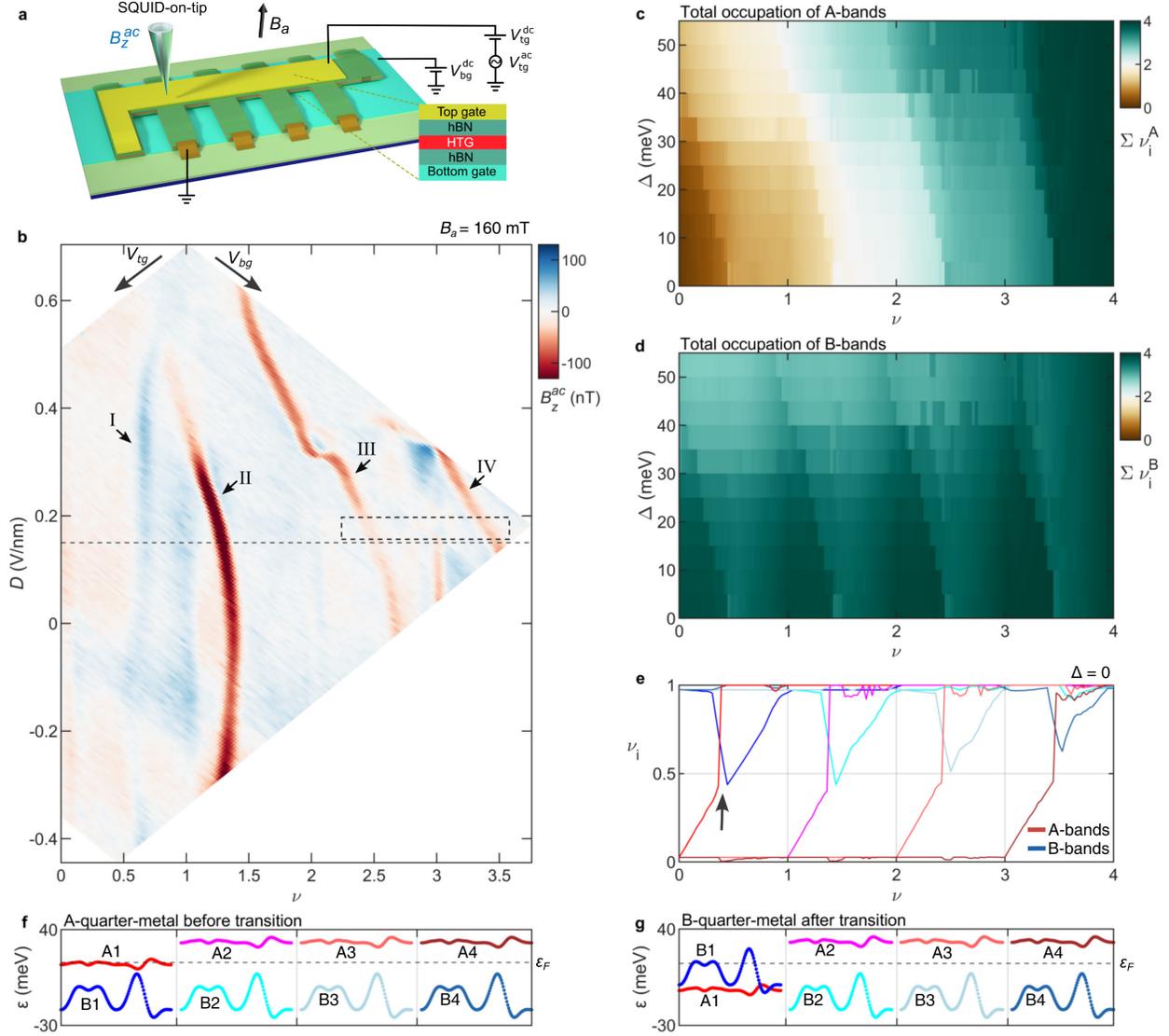

**Fig. 2. Magnetic phase diagram and seesaw sublattice switching. A,** Experimental setup schematics. Top and bottom gate voltages, $V_{tg}^{dc} + V_{tg}^{ac}$ and $V_{bg}^{dc}$ are indicated, along with the corresponding *ac* magnetic field $B_z^{ac}$ imaged by the scanning SOT. **B,** $B_z^{ac}$ measured at a fixed SOT position above device B as a function of $\nu$ and $D$ at $B_a = 160$ mT. The four peak features, labeled I-IV, are robust across the samples (Fig. S2) and various magnetic fields (Fig. S3). **C-D,** The scHF calculation of the total occupation of A- (**C**) and B-bands (**D**) as a function of $\nu$ and displacement potential $\Delta$ for intermediate coupling strength, $\epsilon_r = 11$. The seesaw-like sublattice switching lines follow trajectories similar to experiment (B). **E,** The scHF results of the individual fillings $\nu_i$ of the four A-bands and four B-bands at $\Delta = 0$ for $\epsilon_r = 8$. The bands are color coded as in (**F-G**). **F-G,** Schematics of the sublattice phase transition switching upon doping across the black arrow in (**E**). Before the transition, the system is a metal in an A-band (**F**), whereas after the transition it is a B-band metal (**G**). Blue (red) color shades indicate B (A) polarized bands, while specific shades correspond to different spin-valley flavors labeled 1 to 4.

To gain insight into the underlying mechanism, we employ scHF mean-field theory (Supplementary Materials). In the noninteracting case, the eight low-energy flat bands are not fully gapped (Fig. S6A) and exhibit sublattice mixing (superpositions of A and B sublattices) [14,18]. However, interactions gap out the flat bands, forming two distinct sets: four 'A-bands', highly polarized to the A sublattice with Chern number $C_A = +1 \,(-1)$, and four 'B-bands', predominantly B-sublattice polarized with $C_B = -2 \,(+2)$ in valley $K$ ($K'$) within the h-HTG domain (Fig. S8).

Figures 2C,D show the calculated total occupation of the four A- and four B-bands, $\sum_i \nu_i^{A,B}$, as a function of total $\nu$ and the layer potential difference $\Delta$, controlled by a displacement field $D$. The occupations exhibit a sequence of sharp jumps, with similar trajectories as the phase transition peaks in the differential magnetization (Fig. 2B). At small $\Delta$, these periodic seesaw-like transitions occur within metallic states between integer $\nu$ (Movie S1). As $\Delta$ increases, they shift toward lower densities, eventually crossing integer fillings. These sharp transitions are a robust feature of the HTG phase diagram, consistently observed in both scHF simulations (Fig. S7) and a simplified Stoner model (Fig. S6C), providing strong evidence that they capture the fundamental electronic behavior of HTG.

To understand the seesaw transitions, we analyze the occupations of the eight A- and B-bands at $\Delta = 0$ (Fig. 2E). The four spin-valley flavors, indistinguishable within our scHF calculations (Supplementary Materials), are labeled 1 to 4. At $\nu = 0$, the system starts with four fully occupied B-bands ($\nu_i^B = 1$, blue shades) forming the valence band, and four empty A-bands ($\nu_i^A = 0$, red shades) forming the conduction band. As doping increases, the A1 band (bright red) gradually fills. At a critical filling $\nu \cong 0.4$ (black arrow) a sublattice phase transition occurs: A1 band rapidly fills, while B1 (blue) partially empties, swapping their roles as valence and conduction bands. Before the transition, the system is an A1-quarter-metal (Fig. 2F), which switches to a B1-quarter-metal (Fig. 2G) afterward. As doping continues, the B1-band fills completely at $\nu = 1$. For $\nu > 1$, the process repeats for a different flavor: the A2-band (magenta) fills until, at $\nu \cong 1.4$, another phase transition occurs, where A2 rapidly fills while B2 (cyan) partially empties. This periodic, seesaw-like filling and emptying of A- and B-bands continues until all eight bands are fully occupied at $\nu = 4$.

Although these seesaw transitions involve a substantial reorganization of the electronic structure, they occur entirely within metallic states and therefore produce only weak transport signatures. Because the total carrier density is conserved and the mobilities of the A- and B-derived metallic states are expected to be comparable, the longitudinal resistance changes little across a seesaw transition, even when a feature is discernible (Fig. S2f). Magnetization, by contrast, is a thermodynamic quantity providing a much more sensitive probe of the band reorganization underlying the seesaw transitions, as described below.

The sublattice transitions occur within a single flavor, consistent with the experimentally observed second-order character (Supplementary Materials and Fig. S5). These reversals between valence and conduction bands, differ fundamentally from the symmetry-breaking mechanisms in other interacting flat-band systems [26,33–37] and, as we show next, they give rise to a novel form of hysteresis.

**Local hysteresis**

The hysteresis in $R_{yx}$ near $\nu = 3$ (Figs. 1C,D) indicates that the system transitions between two metastable states related by time reversal (TR) symmetry. These TR transitions likely correspond to switching between valleys $K$ and $K'$, as observed in other Chern-band systems [4–6,9–11,28,31,37,38]. To investigate this further, we measure the hysteresis in the magnetization by sweeping $\nu$ at constant $D$ within the dashed black box marked in Fig. 2B. Figures 3A,B show $B_z^{\mathrm{ac}}(\nu, D)$ measured upon increasing ($\nu_\uparrow$) and decreasing ($\nu_\downarrow$) the carrier density respectively, while Fig. 3C presents their numerical subtraction, $\Delta B_z^{\mathrm{ac}}(\nu) = (B_z^{\mathrm{ac}}(\nu_\uparrow) - B_z^{\mathrm{ac}}(\nu_\downarrow))/2$.

The vertical, $D$-independent feature at $\nu = 3$ (black arrows in Figs. 3A-C) reflects the Chern magnetization in the incompressible Chern gap. This peak has opposite signs for $\nu_\uparrow$ and $\nu_\downarrow$, confirming that the two metastable states correspond to reversed valley populations with Chern insulating states of opposite Chern number.

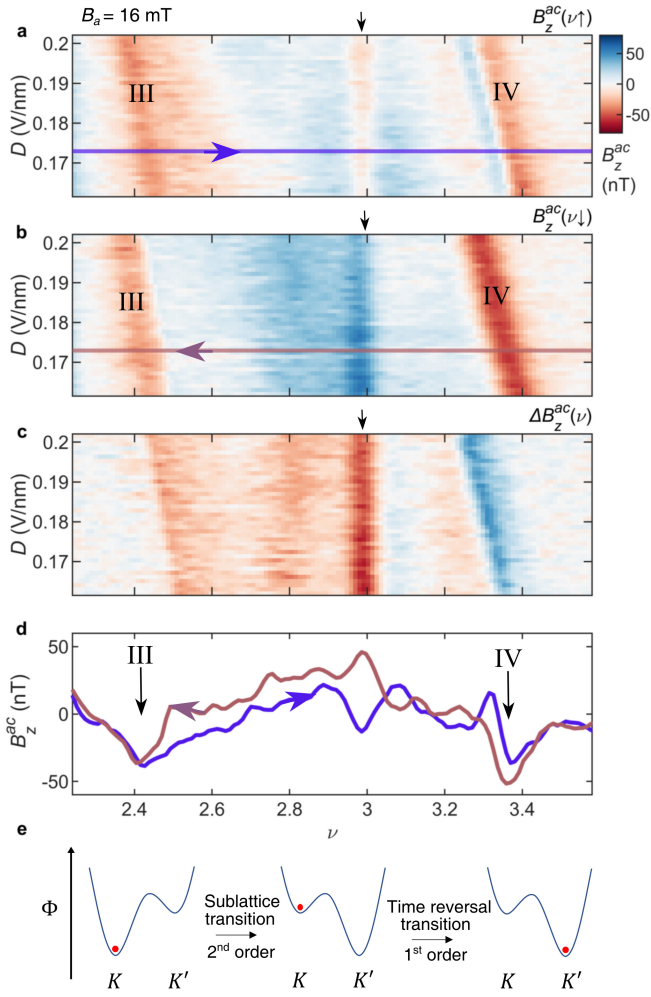

**Fig. 3. Hysteresis in the local magnetization. A-C,** Hysteresis in $B_z^{ac}$ while sweeping $\nu_\uparrow$ (**A**), $\nu_\downarrow$ (**B**), and their numerical difference $\Delta B_z^{ac}(\nu)$ (**C**), vs. $\nu$ and $D$ in the region indicated by the dashed black box in Fig. 2B. Seesaw transition lines III and IV are indicated. The black arrows point to the magnetization peak in the Chern gap at $\nu = 3$, which switches sign between $\nu_\uparrow$ and $\nu_\downarrow$ sweeps. **D,** Linecut of $\nu_\uparrow$ (blue) and $\nu_\downarrow$ (brown) at $D = 0.173$ V/nm, marked by solid lines in (**A,B**). **E,** Schematics of the hysteresis mechanism where the $y$ axis is the free energy $\Phi$ of the system. The sublattice transition raises the energy of $K$ valley triggering a first-order transition into $K'$ valley.

Surprisingly, Fig. 3C reveals that the entire region between the seesaw transitions III and IV is hysteretic, with no hysteresis outside these boundaries. This is further evidenced in the linecut at $D = 0.173$ V/nm (Fig. 3D), where the $\nu_\uparrow$ and $\nu_\downarrow$ trajectories coincide for $\nu < 2.4$ and $\nu > 3.4$, but differ in the entire range of $2.4 < \nu < 3.4$. This behavior contrasts sharply with previous studies of magnetic hysteresis in moiré Chern insulators [4–6,9–11,27,28,31], where hysteresis is confined to the vicinity of the Chern gaps. In those systems, hysteresis arises when the total magnetization of a valley-polarized state reverses sign as described in Ref. [31]—either upon crossing a Chern gap, where the Chern magnetization $M_C$ dominates, or within a metallic state due to a sign change of the self-rotation contribution $M_{SR}$ [27]. In HTG, by contrast, the observed hysteresis is not tied to a magnetization sign change associated with a Chern gap, but is sharply bounded by the two seesaw transitions III and IV, pointing to a fundamentally different origin.

Minor loop measurements (Fig. S10) further elucidate this mechanism. When $\nu$ is swept only within the region preceding seesaw transitions III and IV, no hysteresis is observed. Hysteresis emerges only when the sweep crosses both seesaw transition boundaries, requiring a substantially larger excursion in $\nu$. These measurements reveal that the TR transition between $K$ and $K'$ valleys occurs precisely at peak IV for $\nu_\uparrow$ and at peak III for $\nu_\downarrow$. This indicates that although the seesaw transitions themselves involve sublattice reorganization within a single flavor, they can also trigger a valley-switching TR transition. The schematic in Fig. 3E illustrates this scenario: the sublattice transition raises the energy of one valley, inducing a first-order recondensation of carriers into the opposite valley. Understanding this new mechanism of hysteresis—where TR valley switching is enabled by sublattice-driven electronic reordering—calls for further theoretical investigation.

**Modeling magnetization across seesaw transitions**

We now clarify the connection between the sublattice seesaw transitions and the experimentally observed magnetic peaks by analyzing the orbital magnetization response. We show that while the conduction-band-only null hypothesis fails to reproduce the observed magnetization jumps (Supplementary Materials and Fig. S9), the seesaw transitions capture their key qualitative features. A direct calculation of the orbital magnetization in scHF calculations are both computationally demanding and difficult to interpret physically. Thus, motivated by the scHF finding that interactions in HTG strongly polarize the low-energy bands into A- and B-sublattice sectors [18], we adopt a simplified modeling approach. We compute the single-particle band structure in a basis of sublattice-polarized bands (Fig. 4A,B and Movie S2), which allows for a transparent and direct evaluation of the resulting changes in orbital magnetization (Supplementary Materials).

The orbital magnetization contains two Berry-curvature–induced contributions: the self-rotation magnetization, $M_{SR}$, which depends only on the band structure and its occupation, and the Chern magnetization, $M_C$, which explicitly depends on the chemical potential $\mu$ (Supplementary Materials) [11,29,30]. When filling a metallic state, the shifts in $\mu$ are usually small and limited by the band width, so magnetization changes are dominated by $M_{SR}$. By contrast, when crossing a Chern gap, $M_{SR}$ remains unchanged and $M_C$ accounts for the entire magnetization change, $\Delta M = \Delta M_C = \frac{e}{h} C \Delta \mu$ (here, $e$ is the elemental charge, and $h$ is Planck constant) [11,39].

In HTG, this conventional picture breaks down. Even though the sublattice transitions occur in the metallic states with only a small change in the Fermi level, the flipping of the A- and B-bands gives rise to a drastic change in their individual chemical potentials $\mu_i$ relative to their band minima (see Fig. 4B and Movie S3 for evolution in $K'$ valley). In particular, $\mu_B$ drops sharply by $\delta\mu_B$ (Fig. 4C), which reduces its $M_C$. Simultaneously, $\mu_A$ jumps up by $\delta\mu_A$. However, instead of increasing, its Chern magnetization drops because the Chern number of A-band is negative. As a result, the Chern contributions from the two bands add constructively, $\Delta M_C \simeq C_A \delta\mu_A + C_B \delta\mu_B = -(|C_A \delta\mu_A| + |C_B \delta\mu_B|)$, resulting in a sharp drop in the total magnetization. The corresponding calculated total and differential magnetizations are shown in Figs. 4D,E. The four peaks in the measured differential magnetization (Fig. 4F) are well reproduced qualitatively in the calculations. They reflect the large jumps in the chemical potential of the individual bands that occur in a metallic state without affecting significantly the global Fermi level, in addition to weaker features at integer fillings. This behavior is qualitatively distinct from the conventional case, where the largest changes in $\mu$ and magnetization occur upon crossing gapped states.

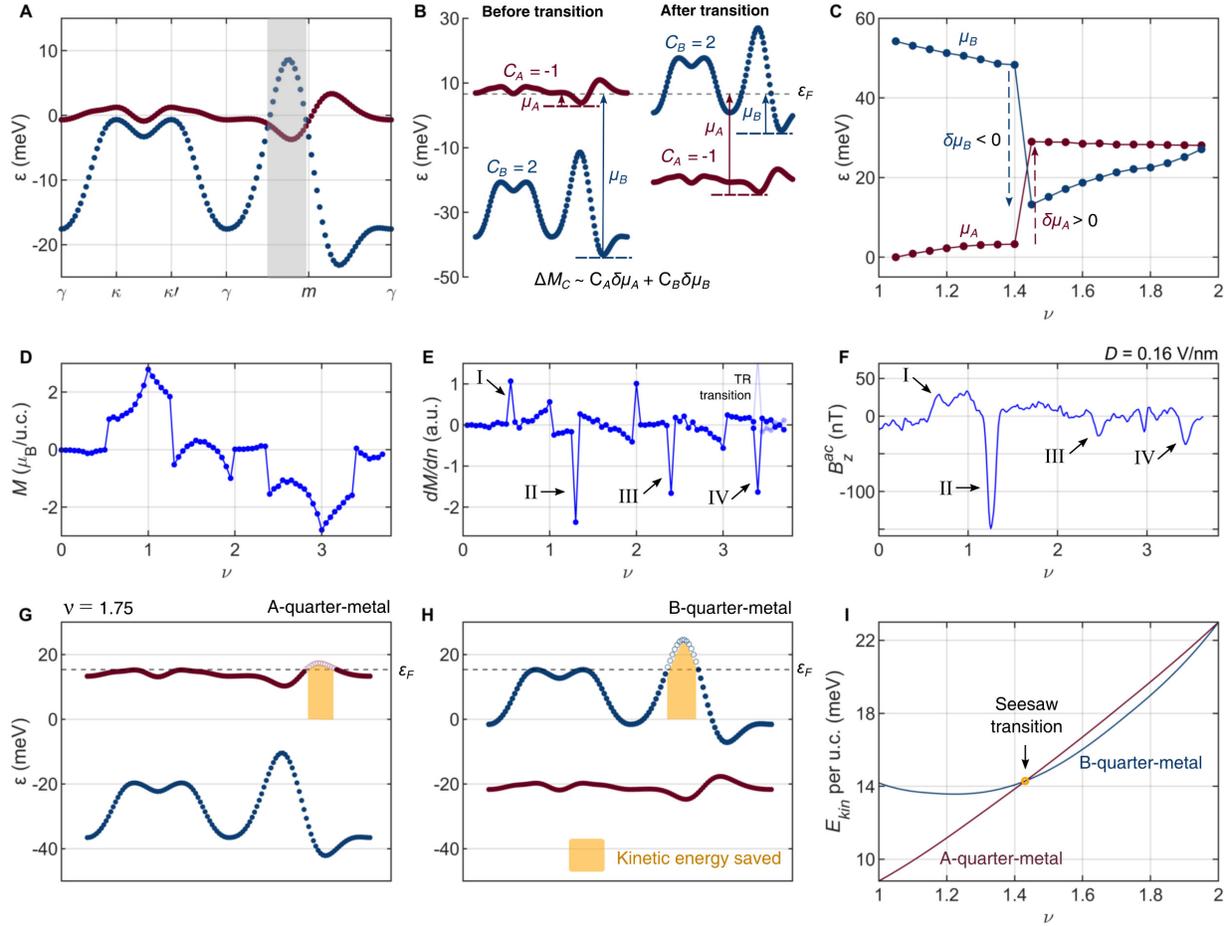

**Fig. 4. Theoretical modeling of magnetization across seesaw transitions. A,** Noninteracting band structure of HTG flat bands with suppressed tunneling between A- and B-bands (Supplementary Materials). Gray strip marks the momentum region with inverted pockets. **B,** Schematic of seesaw transition in $K'$ flavor with $\Delta \varepsilon_F \cong 0$. The individual chemical potentials, $\mu_{A,B}$, are marked. **C,** Evolution of the individual chemical potentials, $\mu_{A,B}$, relative to the band minima as a function of $\nu$ across a seesaw transition at $\nu = 1.4$. A large change in $M_C$ results from the jump in the chemical potentials $\delta\mu_{A,B}$ of individual bands at the transition. **D,** Simulated $M$ as a function of $\nu$ using a model inspired by scHF (Supplementary Materials). **E,** Numerical derivative, $\frac{dM}{dn}$, of (d) with indicated seesaw transitions I to IV. **F,** Linecut of $B_z^{ac}$ along $D = 0.16$ V/nm (Fig. 2B, black dashed line), showing qualitative agreement with simulated $\frac{dM}{dn}$ in (e). **G-H,** Schematic band structures of A- and B-quarter-metal states at $\nu = 1.75$. The kinetic energy gained (orange areas) by emptying the same number of electronic states in B-metal is larger than in A-metal giving rise to flipping of the bands. **I,** Kinetic energy per moiré u.c. as a function of $\nu$ for the two possible ground states. At low doping A-metal is the ground state, while above critical doping (arrow) B-metal has a lower kinetic energy resulting in seesaw transition.

**Discussion**

The symmetry-breaking transitions observed in HTG are fundamentally different from other moiré materials, including MATBG [1,8,11,33,34,40–43], hBN-aligned MATBG [4,6,44], twisted monolayer-bilayer graphene [9,38,45,46], and twisted double bilayer graphene [2,3,47], none of which exhibit a reversal of valence and conduction bands or sublattice polarization upon electron doping. The distinctive features of HTG that allow this unique spin-valley-sublattice tunability can be understood by examining the single-particle dispersion of its sublattice-polarized Chern bands (Fig. 4A). Notably, while the B-band has lower average kinetic energy, it exceeds that of the A-band within a small pocket (Fig. 4A, shaded gray). The total energy of a given ground state is equal to the sum of the interaction energy, $U_{\text{int}} \nu_A \nu_B$, and kinetic energy, $\sum_{A,B} E_{\text{kin}}$. To minimize the interaction energy, the system prefers maximal polarization, favoring either A-metal or B-metal phases (Figs. 4G,H). Since the interaction energy is the same in both, the kinetic energy determines the preferred state. Emptying high-energy states in A-metal gains kinetic energy (shaded area, Fig. 4G), but emptying the same number of states in the more dispersive pocket in B-metal, yields greater kinetic energy gain (Fig. 4H). Figure 4I shows the calculated kinetic energy for the two cases. At low doping, A-metal is favored, but beyond a critical density, B-metal becomes the ground state, leading to the valence-conduction band flip and sublattice transition.

Unlike other graphene-based systems, where interactions choose between two degrees of freedom, spin and valley, here interactions have access to a much larger phase space, manipulating all three degrees of freedom—spin, valley, and sublattice polarization—potentially leading to a new class of symmetry-broken states of matter. Realizing full spin-valley-sublattice tunability through correlations requires flat bands, strong exchange interactions, broken inversion symmetry, and overlapping bands with inverted pockets (shaded gray in Fig. 4A). While many 2D van der Waals systems possess most of these ingredients, the apparent unique feature of HTG is its inverted pockets. Although there is no established method for engineering such pockets, there is no fundamental reason they couldn't emerge in other interacting systems. Moreover, a finely tuned band structure with multiple kinetic energy crossings (Fig. 4I) could even enable more exotic and rapid seesaw band flippings. Since the transitions occur in a compressible state and exhibit only subtle transport signatures, similar transitions may have gone unnoticed in other materials.

Our results establish HTG as a prototypical system exhibiting sublattice seesaw transitions—a new kind of doping-induced phase transitions involving bands with contrasting sublattice polarization. In this regime, the conventional distinction between valence and conduction bands breaks down. Instead, all eight moiré flat bands, characterized by distinct spin, valley and sublattice degrees of freedom, must be considered collectively to minimize the total kinetic and electron-electron Coulomb energy. Beyond uncovering this novel transition, our study highlights the power of magnetic imaging in mapping correlated phase diagrams, particularly in the compressible regime where transport measurements alone often fail to capture key electronic reconstructions. Furthermore, we find that sublattice transitions can directly drive time-reversal transitions between valleys (Supplementary Materials), suggesting an intrinsic link between electronic correlations and orbital magnetism. While a full theoretical understanding of this interplay remains an open question, our findings offer a new perspective on electrical switching in orbital magnets and open new possibilities for designing programmable topological quantum devices.

**Acknowledgments**

**Funding:** This work was co-funded by the Tom and Mary Beck Center for Advanced and Intelligent Materials at the Weizmann Institute of Science, by the Minerva Stiftung with funding from the Federal German Ministry for Education and Research, by the United States - Israel Binational Science Foundation (BSF) grant No 2022013, and by the European Union (ERC, MoireMultiProbe - 101089714). Views and opinions expressed are however those of the author(s) only and do not necessarily reflect those of the European Union or the European Research Council. Neither the European Union nor the granting authority can be held responsible for them. E.Z. acknowledges the support of the Goldfield Family Charitable Trust, the Knell Family Institute for Artificial Intelligence, and Leona M. and Harry B. Helmsley Charitable Trust grant #2112-04911. This work was partially supported by the Army Research Office MURI (grant no. W911NF2120147), the 2DMAGIC MURI (grant no. FA9550-19-1-0390), the National Science Foundation (grant no. DMR-1809802), the Office of Naval Research (grant no. N000142412440), the Ramón Areces Foundation and the Gordon and Betty Moore Foundation's EPiQS Initiative through grant no. GBMF9463 to P.J.H. K.W. and T.T. acknowledge support from the JSPS KAKENHI (Grant Numbers 21H05233 and 23H02052), the CREST (JPMJCR24A5), JST and World Premier International Research Center Initiative (WPI), MEXT, Japan. M.B. acknowledges the VATAT Outstanding PhD Fellowship in Quantum Science and Technology. A.U. acknowledges support from the MIT Pappalardo Fellowship and from the VATAT Outstanding Postdoctoral Fellowship in Quantum Science and Technology. L.-Q.X. acknowledges support from the MathWorks Fellowship. A.L.S. was supported by the US Department of Energy, Office of Science, Basic Energy Sciences, Materials Sciences and Engineering Division, under Contract DE-AC02-76SF00515. T.D. acknowledges support from a startup fund at Stanford University.

**Author contributions:** I.R., W.Z., M.B. and E.Z. designed the experiment. I.R., W.Z., and M.B. performed the measurements. L.-Q.X. and A.U. designed and fabricated the samples under the supervision of P.J.-H., with help from M.U. The SOTs were fabricated by I.R. and W.Z., and Y.M. fabricated the tuning forks. T.D. and Y.H.K. performed band structure, magnetization, and Hartree–Fock calculations. K.W. and T.T. provided the hBN crystals. M.B., I.R., W.Z., E.Z., L-Q.X., A.U., A.S., T.D. and Y.H.K contributed to data analysis and theoretical modeling. M.B., W.Z., L-Q.X., A.U., Y.H.K, and E.Z. wrote the original manuscript. All authors participated in discussions and revisions of the manuscript.

**Competing interests** The authors declare no competing interests.

**Data availability** The data that support the findings of this study are available from the corresponding authors on reasonable request.

**Code availability** scHF calculations used in this study are available from the corresponding authors on reasonable request.

**Supplementary Materials**
Materials and Methods
Supplementary Text
Figs. S1 to S10
References (*49-54*)
Movies S1 to S3

# Supplementary Materials for

**Electrically controllable valence-conduction band reversals in helical trilayer graphene**

**Authors:** Matan Bocarsly[1†], Indranil Roy[1†], Weifeng Zhi[1†], Li-Qiao Xia[2†], Aviram Uri[2], Yves H. Kwan[3], Aaron Sharpe[4,5], Matan Uzan[1], Yuri Myasoedov[1], Kenji Watanabe[6], Takashi Taniguchi[7], Trithep Devakul[4], Pablo Jarillo-Herrero[2]* and Eli Zeldov[1]*

e-mail: pjarillo@mit.edu, eli.zeldov@weizmann.ac.il

**The PDF file includes:**

    Materials and Methods
    Supplementary Text
    Figs. S1 to S10
    References (*49-54*)
    Movies S1 to S3

**Other Supplementary Materials for this manuscript include the following:**

    Movies S1 to S3

## Materials and Methods

### Device fabrication

The van der Waals heterostructures were assembled in two parts using the standard dry-transfer technique. For graphite backgate (Device A), an hBN flake and a few-layer graphene strip were picked up by a poly(bisphenol A carbonate) stamp. The bottom stack was released onto a 285 nm SiO$_2$/Si substrate. For metal backgate (Devices B and C), only an hBN flake was picked up and released onto the pre-deposited metallic strip (16 nm Au/Pd 60/40 weight percent with 2 nm Ti as the adhesion layer). After dissolving the stamp in chloroform, the bottom stack was annealed at 350 °C in vacuum for 12 hours to remove polymer residues. Then, tip cleaning was performed using the contact mode of a Bruker Icon XR atomic force microscope to clean the surface further. The twisted graphene stack was made following the standard cut-and-stack procedure. A monolayer graphene flake was cut into three pieces using a confocal laser-cut setup. A second poly(bisphenol A carbonate) stamp was used to pick up an hBN flake and the three graphene pieces subsequently. Before picking up the second and third pieces of graphene, the stage was rotated by 1.8° in the same direction to realize the helical stacking order. The pickup of graphene was done at room temperature to avoid spontaneous changes in the orientations of the graphene layers. The top stack was released onto the bottom stack at temperatures between 150 and 170 °C.

The Hall bars were defined in bubble-free regions identified under an atomic force microscope. Patterning was performed using an Elionix ELS-HS50 electron-beam lithography system. Metallic top gates (21 to 65 nm Au with 2 to 5 nm Cr or Ti adhesion layers) were deposited using a Sharon thermal evaporator. The devices were connected using one-dimensional contacts (40 to 63 nm Au with 2 to 5 nm Cr adhesion layer). Finally, the devices were etched into a Hall bar geometry using reactive-ion etching. All devices had twist angles close to the magic angle, $\theta_\mathrm{m} \approx 1.8°$.

Device summary:

**Device A**: Graphite bottom gate, Ti/Au top gate; bottom hBN thickness ≈ 44 nm, bottom gate capacitance $C_\mathrm{bg} \approx 3.77\times10^{11}$ e·cm$^{-2}$·V$^{-1}$, top hBN thickness ≈ 14.2 nm, top gate capacitance $C_\mathrm{tg} \approx 1.18\times10^{12}$ e·cm$^{-2}$·V$^{-1}$.

**Device B** (local measurements presented in main text): Ti/Au bottom gate, Ti/Au top gate; bottom hBN thickness ≈ 14.4 nm, $C_\mathrm{bg} \approx 1.06\times10^{12}$ e·cm$^{-2}$·V$^{-1}$, top hBN thickness ≈ 22.7 nm, $C_\mathrm{tg} \approx 6.7\times10^{11}$ e·cm$^{-2}$·V$^{-1}$.

**Device C**: Ti/Au bottom gate, Ti/Au top gate; bottom hBN thickness ≈ 14.4 nm, $C_\mathrm{bg} \approx 1.06\times10^{12}$ e·cm$^{-2}$·V$^{-1}$, top hBN thickness ≈ 22.7 nm, $C_\mathrm{tg} \approx 6.7\times10^{11}$ e·cm$^{-2}$·V$^{-1}$.

### Transport measurements

Four-point transport measurements were performed at $T = 300$ mK using standard lock-in techniques with a bias current of $I = 10$ nA rms at 11 Hz. Longitudinal and transverse resistances $R_{xx}$ and $R_{yx}$ are acquired as a function of the top and bottom gate *dc* voltages, $V_\mathrm{tg}^\mathrm{dc}$ and $V_\mathrm{bg}^\mathrm{dc}$. The data was then plotted as a function of $n = C_\mathrm{tg}V_\mathrm{tg}^\mathrm{dc} + C_\mathrm{bg}V_\mathrm{bg}^\mathrm{dc}$ and $D = \left(C_\mathrm{tg}V_\mathrm{tg}^\mathrm{dc} - C_\mathrm{bg}V_\mathrm{bg}^\mathrm{dc}\right)/2\epsilon_0$, with capacitances determined from fitting of the Landau fan diagrams ($\epsilon_0$ is the permittivity of free space). All devices showed a twist angle of $\theta_\mathrm{m} \approx \sqrt{\frac{\sqrt{3}}{8}a^2 n_\mathrm{s}} \approx 1.8°$ ($a = 0.246$ nm), determined from the measurement of the carrier density $n_\mathrm{s}$, corresponding to four electrons per moiré unit cell.

**SOT fabrication and local magnetization measurements**

An Indium SOT of 180 nm diameter was used to measure the *ac* magnetic stray field emanating from the sample, $B_z^{ac}$. The SOT was fabricated as described in [32,48,49]. The SOT was attached to quartz tuning fork excited at its resonance frequency of $\approx$ 33 kHz for sample approach and height control as described in [50]. The magnetic imaging was acquired at constant height of about 180 nm above the sample surface. The measurements were performed at $T = 300$ mK using a cryogenic SQUID series array amplifier (SSAA) [51]. The SOT had a magnetic field sensitivity down to 10 nT/Hz$^{1/2}$. Owing to the interference pattern of the SOT, its magnetic sensitivity is minimal near zero applied field and maximal near half-integer flux quanta threading the loop. A small external magnetic field is therefore applied to bias the SOT at its optimal operating point.

The $B_z^{ac}$ images were obtained with pixel size of 90 nm and acquisition time of 1 s/pixel. The measured signal $B_Z^{ac}$ is the response to a small *ac* voltage $V_{tg}^{ac}$ of typically 50 mV rms applied to the top gate at a frequency of $f \approx 6$ kHz. Since the top gate voltage modifies both $n$ and $D$, the signal can be written as $B_z^{ac} = n^{ac}(\partial B_z/\partial n) + D^{ac}(\partial B_z/\partial D)$, where $n^{ac}$ and $D^{ac}$ are the corresponding modulations induced by the applied $V_{tg}^{ac}$, with the dominant contribution arising from $n^{ac}$ modulation. The amplitude of $V_{tg}^{ac}$ was optimized for best signal to noise ratio, while avoiding smearing of the signal.

**Supplementary Text**

**Domain structure and numerical reconstruction of differential magnetization**

Theory predicts formation of h-HTG and $\bar{\text{h}}$-HTG spatial domains with opposite Chern numbers per valley [14], which should lead to magnetic domains with opposite magnetization. Figure S1a shows an area scan of device C with stripe-like domains of alternating magnetic signal consistent with such a domain structure. In contrast, devices A and B (Figs. S1b,c) show no evidence of spatial domains, exhibiting a magnetic response that is uniform in sign over the scanned area. It has been shown theoretically that weak heterostrain can greatly enlarge the domain sizes [20], potentially beyond the scan area, which may explain the apparent single-domain behavior in devices A and B. Consistent with this, experimental studies have demonstrated that domain sizes can vary significantly, for example following thermal cycling [20]. Importantly, our analysis and theoretical modeling address the electronic phases within an individual h-HTG or $\bar{\text{h}}$-HTG domain. Device B, presented in the main text, realizes this single-domain regime over the full scanned area.

Since $B_z^{ac}$ signal reflects the stray field generated by the *ac* modulation in the local magnetization, the 2D map of $B_z^{ac}(x,y)$ allows direct numerical reconstruction of the local differential magnetization $m_z(x,y) = dM_z(x,y)/dn$ [52]. Figures S1c,d show an example of $B_z^{ac}(x,y)$ and the corresponding reconstructed $m_z(x,y)$ at $\nu = 1.30$ in device B. A sharp negative peak (red) is observed in the central and left parts of the sample, showing exceptionally high magnetization values reaching 30 $\mu_\text{B}$ per electron added to the system. As shown below, the integrated magnetization changes are far too large to arise from spin polarization and instead signal abrupt changes in orbital magnetization associated with interaction-driven electronic reordering. In conventional symmetry-broken topological bands, the largest magnitude of the differential orbital magnetization is expected to occur in the Chern gaps, as observed in MATBG [10,11]. Here the behavior is opposite, with the strongest magnetism appearing in the metallic state at $\nu = 1.30$ with much weaker magnetic response at integer fillings (Fig. 2B and Figs. S2 and S3). This inverted hierarchy directly reflects the large change in Chern magnetization induced by the sublattice seesaw transition.

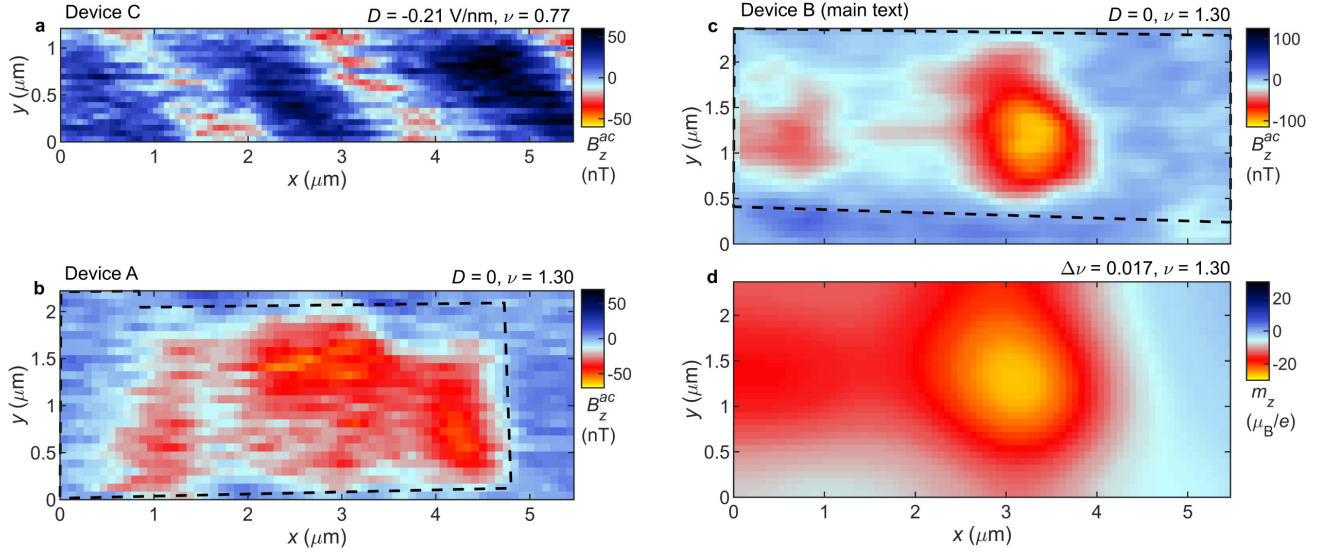

**Fig. S1. Area scans and magnetization reconstruction. a,** Measured 2D map $B_z^{ac}(x,y)$ at $\nu = 0.77$, $D = 0.21$ V/nm, and $B_a = 20$ mT in device C. Stripe-like features of alternating positive and negative signal likely correspond to h-HTG and $\bar{\text{h}}$-HTG domains. **b,** 2D map of $B_z^{ac}(x,y)$ at $\nu = 1.30$, $D = 0$, and $B_a = 34$ mT in device A, where no HTG domains are observed. Dashed lines mark the edge of the sample. **c,** Map of $B_z^{ac}(x,y)$ at $\nu = 1.30$, $D = 0$, and $B_a = 20$ mT in device B (main text), with no observed HTG domains. The $(n,D)$ points at which the images a-c were acquired are marked with magenta dots in Fig. S2. **d,** Differential magnetization $m_z(x,y) = dM_z(x,y)/dn$ reconstructed from (b), with magnitude reaching 30 $\mu_B$/electron.

## Comparison of local $B_z^{ac}$ between different samples and locations

We performed $D$ vs. $\nu$ maps of $B_z^{ac}$ in the three different devices, and at different positions in device B as shown in Fig. S2. In all the measurements, the four seesaw transition peaks can be discerned, albeit with some peaks more pronounced than others in the various measurements. Since the magnetic phase diagrams are consistent, we use the four peaks as the basis for our theoretical explanation of the physics at play in HTG, as described in the main text and below.

Figure S3 shows $B_z^{ac}(\nu, D)$ maps acquired in device B at the same position at various $B_a$ ranging from 17 mT to 259 mT. For $B_a > 100$ mT, the magnetic phase diagram remains unchanged. At lower fields, in addition to the four magnetic peaks labeled I-IV in the main text, additional peaks are resolved, mainly in the vicinity of $\nu = 2$. At our lowest $B_a = 17$ mT (Fig. S3a), the magnetic phase diagram appears to be richer and more complicated, which we cannot capture with our simplified model. Surprisingly, there is a magnetic peak at $\nu = 2$ over a large range of $D$, suggesting a Chern gap. However, the absence of AHE at $\nu = 2$ indicates a trivial gap. One possible explanation is that a spin transition occurs at $\nu = 2$ and that the magnetic peak reflects spin magnetization rather than orbital magnetization. However, more theoretical and experimental work is required for better understanding of the low $B_a$ phases.

## Integration of $B_z^{ac}$

To attain an estimate of the full stray magnetic field $B_z$ as a function of $D$ and $\nu$, we perform higher resolution measurements over a smaller region of phase space. Since the small *ac* modulation in density, $n^{ac} = C_{tg} V_{tg}^{ac}$, is induced by top gate $V_{tg}^{ac}$, we integrate $B_z^{ac}$ over the top gate axis such that $B_z = C_{tg} \int (B_z^{ac}/n^{ac}) \, dV_{tg}$.

Figure S4a shows the raw data as a function of $V_{bg}$ and $V_{tg}$ in device B, while Fig. S4b presents the integrated data. Magnetic peaks I and II are labeled. Line cuts of raw and integrated data at constant $D = -0.2$ V/nm are presented in Figs. S4c,d. The integrated $B_z$ remains negligible until magnetic peak I ($\nu \cong 0.5$) where there is a small positive jump. At magnetic peak II ($\nu \cong 1.3$), $B_z$ has a large negative jump with magnitude $\Delta B_z \cong$ 500 nT. Very similar behavior is observed in device A (Figs. S4f-h).

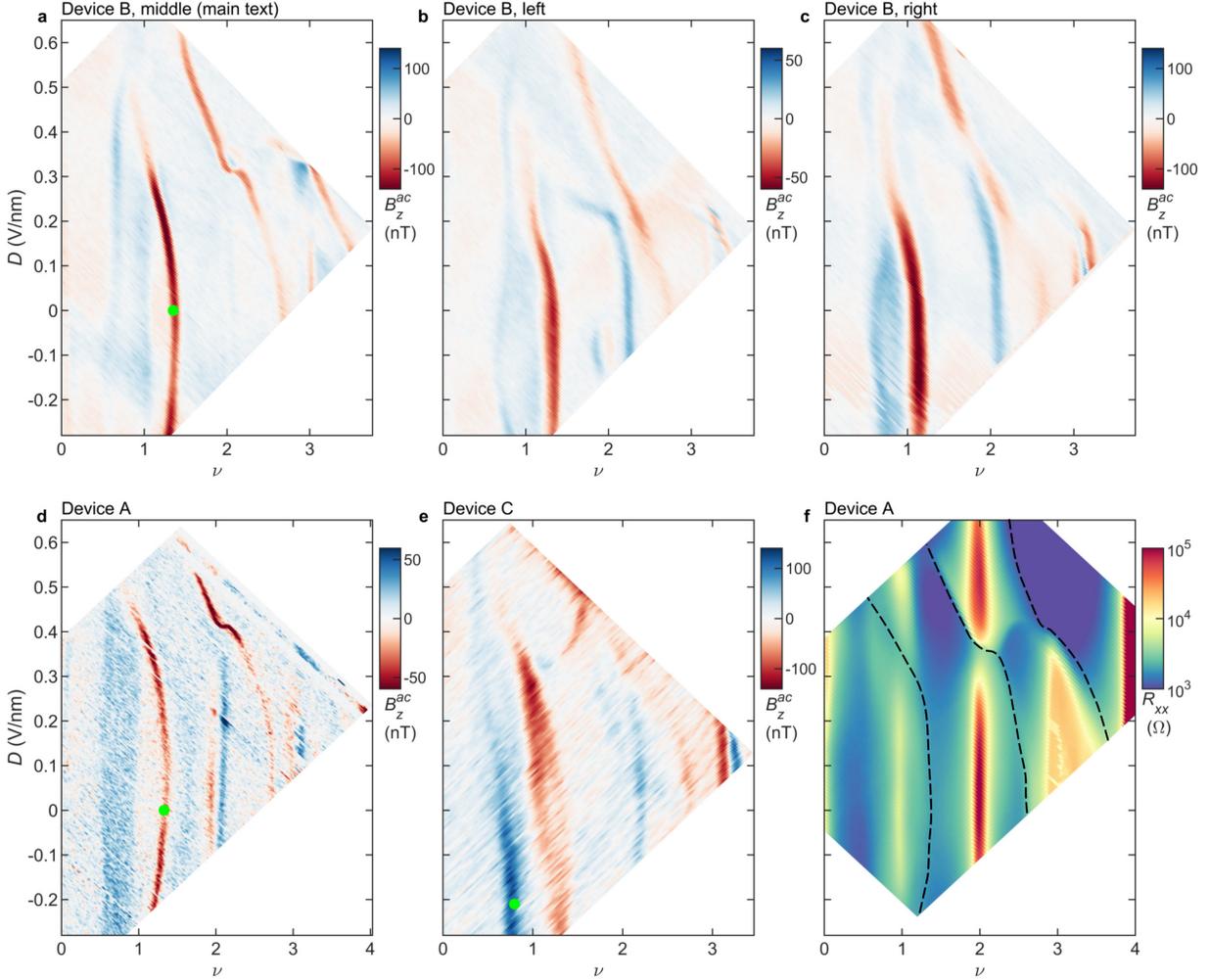

**Fig. S2. Comparison of magnetic phase diagram across samples. a-e,** $B_z^{ac}$ measured at a fixed SOT positions above different samples as a function of $\nu$ and $D$. **a,** $B_z^{ac}$ measured in the center of device B, reproduced from Fig. 2B. **b,c,** $B_z^{ac}$ measured in the left and right parts of device B. **d,** $B_z^{ac}(\nu, D)$ map in device A. **e,** $B_z^{ac}(\nu, D)$ map in device C. The four seesaw transition peak features are robust across different positions and samples. Green dots mark the $(n,D)$ points at which the images in Fig. S1 were acquired. **f,** $R_{xx}$ vs. $\nu$ and $D$ in device A reproduced from Fig. 1A. Resistance minima that follow similar trajectories as the magnetic peaks are highlighted by dashed lines.

It would be instructive to examine the integrated signal over the full range of $\nu$. However, the measured $B_z^{ac}$ contains a very small background contribution arising from noise, which accumulates upon integration and produces a growing offset. As a result, the value of the integrated signal becomes unreliable when integrated over large ranges of $\nu$, and only relative changes in the integrated field, $\Delta B_z$, can be quantified with confidence over limited intervals. Accordingly, we integrate the data shown in Fig. S3a along the line $V_{bg} = 2.5$ V, but present $\Delta B_z$ only over narrow ranges of $\nu$ in the vicinity of the magnetic peaks (Fig. S4e). Each integrated $\nu$ interval is offset to zero, emphasizing that only changes in the integrated signal are physically meaningful.

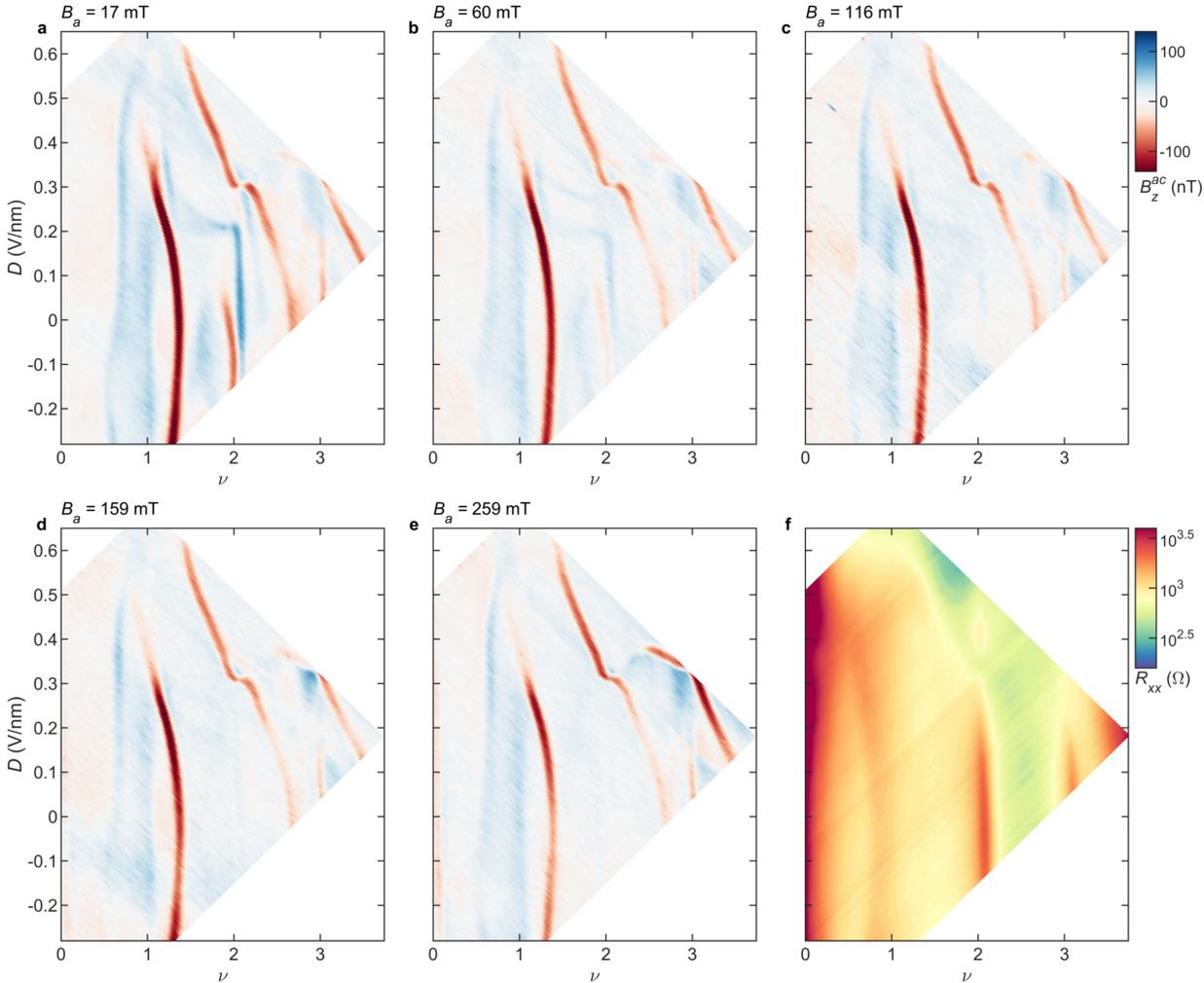

**Fig. S3. Magnetic field dependence of the phase diagram. a-e,** $B_z^{ac}(\nu, D)$ measured at a fixed SOT position in device B (main text) at different indicated external applied magnetic fields, $B_a$. At low $B_a$ additional magnetic features are observed. **f,** $R_{xx}$ vs. $\nu$ and $D$ in device B at $B_a = 0$ mT. Weak $R_{xx}$ features and minima are discerned along trajectories similar to the magnetic peak lines.

To estimate the value of the corresponding jump in magnetization $\Delta M_z$, we simulate a disc of 1 $\mu$m diameter with a constant magnetization. The resulting spatial maps of $B_z$ under experimental conditions and of $M_z$ are shown in Figs. S4i,j. Note that in two dimensions, magnetization and current have the same units, so a magnetic domain can be thought of as a circulating current at its edges. Linecuts in Figs. S4k,l at $y = 0$ show that a domain of this size has approximately constant $B_z$ and $M_z$ within the domain, and that $M_z$ of 1 $\mu_B$ per moiré unit cell results in $B_z \approx 175$ nT. This conversion ratio is used for the right $y$ axis in Figs. S4d,h. Therefore, the measured $\Delta B_z \approx 500$ nT at magnetic peak II corresponds to a jump in magnetization of about $\Delta M_z \cong 3$ $\mu_B$/u.c. Such a large $\Delta M_z$ change over such a small $\Delta \nu \cong 0.15$ implies that the observed magnetization peak features are phase transitions due to electronic reordering in the metallic state. Note that the simple correspondence between $B_z$ and $M_z$ is valid when the magnetic domain size is larger than the SOT diameter. This criterion is met in devices A and B (Fig. S1b,c), but not in device C (presented only in Fig. S1a).

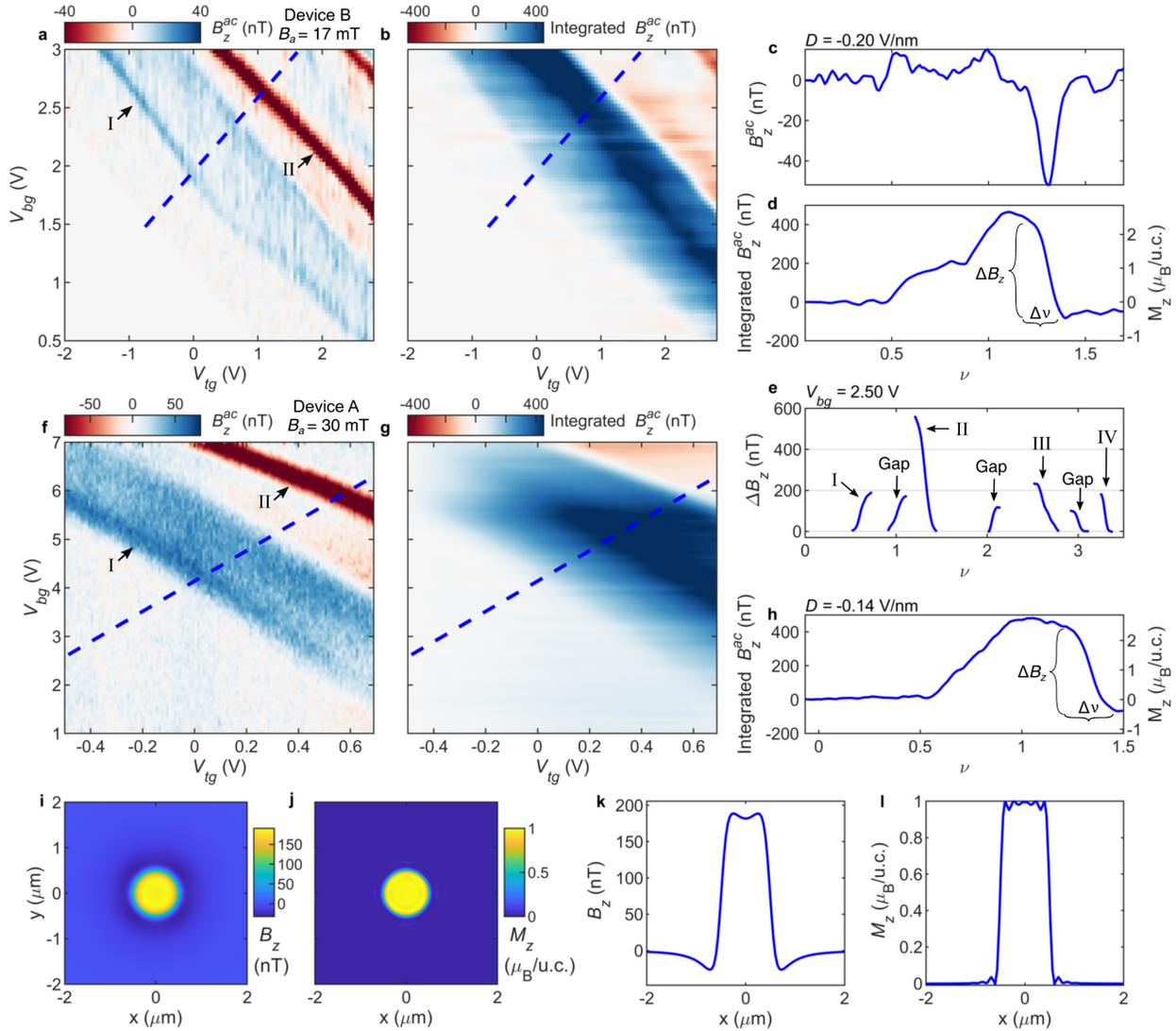

**Fig. S4. Integration of $B_z^{ac}$. a,** Higher resolution $B_z^{ac}$ map over small region of $V_{bg}$ and $V_{tg}$ where the magnetic peaks I and II are discerned in device B (main text). **b,** Integration of $B_z^{ac}$ in (a), $B_z = C_{tg} \int (B_z^{ac}/n^{ac}) \, dV_{tg}$, along $V_{tg}$. **c,** Linecut at constant $D$ of $B_z^{ac}$ along the dashed line in (a). **d,** Linecut at constant $D$ of integrated $B_z$ along the dashed line in (b). Right $y$ axis corresponds to approximate conversion of $B_z$ to $M_z$ derived from simulations shown in (i-l). **e,** Integration of the $B_z^{ac}$ data shown in Fig. S3a along the line $V_{bg} = 2.5$ V. Only regions near the magnetic peaks are shown; each curve is offset to zero and plotted as $\Delta B_z$. **f-h,** Same as (a,b,d) for device A. **i-j,** Numerical simulation of stray magnetic field (**i**) resulting from a circular domain with magnetization of $1 \, \mu_B/u.c.$ (**j**), using the scan height and SOT diameter corresponding to our experimental conditions. **k,l**, Linecuts of (I,j) along $y = 0$, showing that a disc of 1 μm diameter with $1 \, \mu_B/$u.c. magnetization results in approximately constant stray field of $B_z \approx 175$ nT within the disc area.

Figure S4e shows that the magnetization steps associated with the metallic transitions I–IV are systematically larger than those observed at the incompressible gaps. The quantitative results of the integrated peaks are summarized in Table 1. For each transition, we estimate the maximum magnetization change expected from a purely spin-driven transition, $\Delta M_{spin}$, based on the corresponding carrier density $\nu_{transition}$, assuming transitions between fully spin-polarized and fully spin-unpolarized states. The integrated magnetic signal $\Delta B_z$ is extracted from Fig. S4e, and converted to the corresponding magnetization change $\Delta M_z$ using the procedure described above. For peaks I, II, and IV, the extracted $\Delta M_z$ exceeds the maximum possible $\Delta M_{spin}$, ruling out a purely spin origin and establishing these transitions as changes in orbital magnetization. Although peak III could, in principle, be compatible with a spin transition based on its magnitude alone, its similar trajectory in the $\nu - D$ phase diagram strongly suggests a common origin with the other peaks, and therefore an orbital character as well. We also estimate the associated gap sizes by analyzing the magnetization steps associated with integer filling factors. We use the extracted $\Delta B_z$ and corresponding $\Delta M_z$ values (Fig. S4e), and assume $|C| = 1$ at $\nu = 1$ and 3 (corresponding to one fully filled or emptied A band). The resulting gaps, of order a few meV, are significantly smaller than those predicted by the strong-coupling scHF picture discussed below, but are comparable to experimentally extracted gaps in other flat-band systems [53]. At $\nu = 2$, the absence of an anomalous Hall effect indicates a topologically trivial gap, making the origin of the corresponding magnetization step less clear.

| Seesaw transitions | $\Delta B_z$ (nT) | $\Delta M_z$ (μ$_B$/u.c.) | $\nu_{transition}$ | $\Delta M_{spin}$ (μ$_B$/u.c.) |
|---|---|---|---|---|
| Peak I | 190 | 1.1 | 0.6 | 0.6 |
| Peak II | 560 | 3.2 | 1.3 | 1.3 |
| Peak III | 230 | 1.3 | 2.7 | 1.3 |
| Peak IV | 180 | 1.0 | 3.3 | 0.7 |
| | | | | |
| **Gap transitions** | $\Delta B_z$ (nT) | $\Delta M_z$ (μ$_B$/u.c.) | $|C|$ | **Gap** (meV) |
| $\nu = 1$ | 170 | 1.0 | 1 | 4.4 |
| $\nu = 2$ | 110 | 0.6 | 0 | -- |
| $\nu = 3$ | 100 | 0.6 | 1 | 2.6 |

**Table 1**. **Integrated magnetic response across metallic seesaw transitions (I – IV) and integer filling gaps.** Listed are the integrated stray-field change $\Delta B_z$, the corresponding magnetization step $\Delta M_z$, the filling factor at the transition $\nu_{transition}$, and the maximum magnetization change expected from a spin transition $\Delta M_{spin}$. For gap transitions, the assumed Chern number $|C|$ and extracted gap size are shown.

**Determination of second-order seesaw phase transitions**

We checked the dependence of $B_z^{ac}$ on the excitation amplitude $V_{tg}^{ac}$ across the different seesaw transitions. Figures S5a,c show examples of $B_z^{ac}$ measured across magnetic peak II in devices A and B. Linecuts along the dashed vertical lines are presented in Figs. S5b,d, showing linear dependence of $B_z^{ac}$ on $V_{tg}^{ac}$ extending all the way down to zero excitation amplitude. This dependence is characteristic of a second-order phase transition with a continuous $M_z(\nu)$ evolution. In contrast, for a first-order phase transition with a discontinuous step-like jump in $M_z(\nu)$, the $B_z^{ac}$ signal should be zero below some threshold value of $V_{tg}^{ac}$ determined by the size of the hysteresis at the first-order transition and constant above this threshold independent of $V_{tg}^{ac}$. All the seesaw transition magnetic peaks showed similar linear dependence indicating second-order transitions. This suggests that the seesaw transitions exchange roles of valence and conduction bands within a single flavor, since a transition that also changed flavor occupation would be first order, as described in the scHF section below.

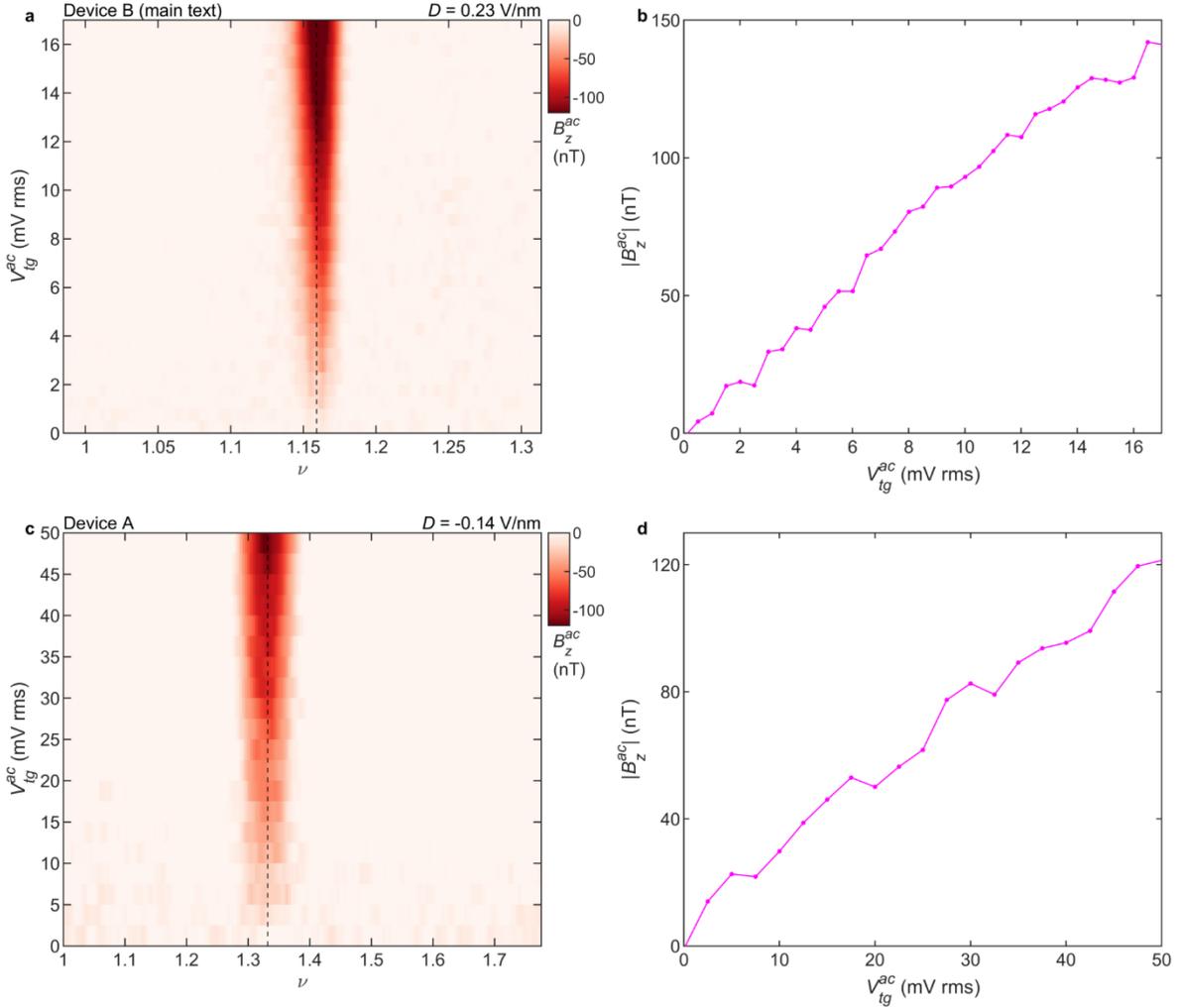

**Fig. S5. Second-order seesaw phase transitions. a,** $B_z^{ac}$ vs. $V_{tg}^{ac}$ excitation amplitude upon sweeping $\nu$ across seesaw magnetic peak II in Device B (main text). **b,** Linecut along the vertical dashed line in (a), showing a linear dependence of $B_z^{ac}$ on the ac excitation amplitude, characteristic of a second-order phase transition. **c-d,** Same as (a-b) in device A consistent with second-order phase transition. A second-order phase transition is most consistent with seesaw transitions that switch between valence and conduction bands within a single flavor.

**Noninteracting continuum model and Chern-sublattice basis**

The non-interacting moiré-periodic band structure in one of the domains of HTG can be described by a generalized Bistritzer-MacDonald continuum model. In the absence of a displacement field, the Hamiltonian for the h-HTG domain in valley $K$ reads

$$H = \begin{pmatrix} -iv_F \boldsymbol{\sigma} \cdot \boldsymbol{\nabla} & T(\boldsymbol{r} - \boldsymbol{d}_t) & 0 \\ T^\dagger(\boldsymbol{r} - \boldsymbol{d}_t) & -iv_F \boldsymbol{\sigma} \cdot \boldsymbol{\nabla} & T(\boldsymbol{r} - \boldsymbol{d}_b) \\ 0 & T^\dagger(\boldsymbol{r} - \boldsymbol{d}_b) & -iv_F \boldsymbol{\sigma} \cdot \boldsymbol{\nabla} \end{pmatrix}$$

where the 2 × 2 blocks act in microscopic sublattice space $\sigma = A, B$, the blocks are ordered according to layer $l = 1,2,3$, and $\boldsymbol{\sigma} = (\sigma_x, \sigma_y)$. The Hamiltonian for valley $K'$ can be obtained by time-reversal symmetry. The diagonal blocks generate the graphene Dirac cones with Fermi velocity $v_F = 8.8 \times 10^5$ ms$^{-1}$, and we have neglected the small Pauli twists due to $\theta$. Note that we have expressed the Hamiltonian in a layer-dependent gauge where $\boldsymbol{k} = 0$ is centered at the Dirac momentum of that layer. The Dirac momenta of the three layers lie at $(\boldsymbol{K}_1, \boldsymbol{K}_2, \boldsymbol{K}_3) = \boldsymbol{K}_D \hat{x} + (k_\theta, 0, -k_\theta) \hat{y}$, where $K_D = \frac{4\pi}{3a_0}$, $k_\theta = K_D \sin \theta$, and $a_0 = 0.246$ nm is the graphene lattice constant. The $\kappa, \gamma, \kappa'$ high-symmetry points of the mBZ are chosen to coincide with $\boldsymbol{K}_1, \boldsymbol{K}_2, \boldsymbol{K}_3$ respectively.

The off-diagonal blocks $T(\boldsymbol{r})$ of the continuum Hamiltonian capture the interlayer tunneling processes. The relative interlayer moiré shifts that characterize the h-HTG domain are given by $\boldsymbol{d}_t - \boldsymbol{d}_b = \frac{1}{3}(\boldsymbol{a}_2 - \boldsymbol{a}_1)$, with $\boldsymbol{a}_{1,2} = \frac{4\pi}{3k_\theta}(\pm\frac{\sqrt{3}}{2}, \frac{1}{2})$ the moiré lattice vectors. The tunneling matrices take the form

$$T(\boldsymbol{r}) = \begin{pmatrix} w_{AA} t_0(\boldsymbol{r}) & w_{AB} t_{-1}(\boldsymbol{r}) \\ w_{AB} t_1(\boldsymbol{r}) & w_{AA} t_0(\boldsymbol{r}) \end{pmatrix}$$

$$t_\alpha(\boldsymbol{r}) = \sum_{n=0,1,2} e^{\frac{2\pi i n \alpha}{3} + i \boldsymbol{q}_n \cdot \boldsymbol{r}} (1 + \lambda_{MDT} \hat{\boldsymbol{q}}_{n,\perp} \cdot \boldsymbol{\nabla})$$

where $\boldsymbol{q}_1 = k_\theta(\frac{1}{2}, \frac{\sqrt{3}}{2})$, $\boldsymbol{q}_{j+1} = R\left(\frac{2\pi}{3}\right) \boldsymbol{q}_j$, and $k_\theta \hat{\boldsymbol{q}}_{j,\perp} = R\left(\frac{\pi}{2}\right) \boldsymbol{q}_j$. We use the tunneling parameters $w_{AA} = 75$ meV, $w_{AB} = 110$ meV, and $\lambda_{MDT} = -2.3$ Å. The latter parameter captures the effects of longer-range interlayer tunneling processes, which is crucial for explaining the pronounced particle-hole asymmetry in this system. An example band structure is shown in Fig. S6a, which exhibits a pair of narrow bands at charge neutrality that are separated from remote bands by large gaps.

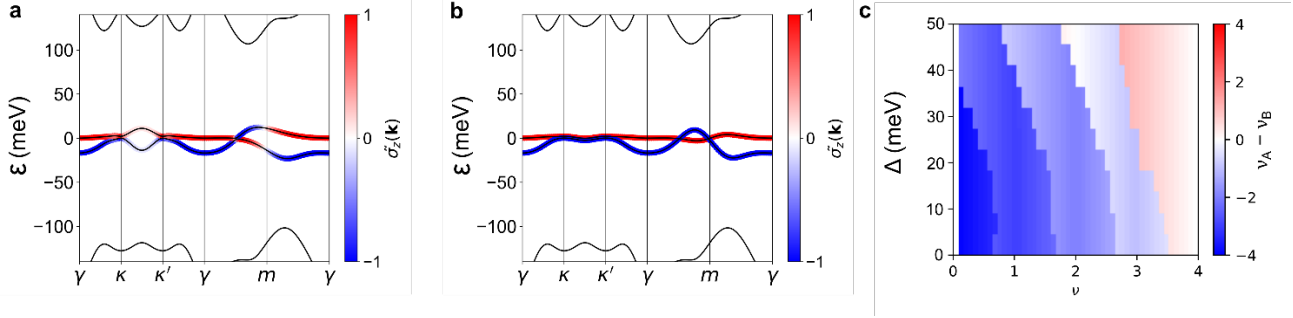

**Fig. S6. Noninteracting band structure of HTG. a,** Band structure at $\theta = 1.8°$ in valley K. For the flat bands, $\tilde{\sigma}_z(\boldsymbol{k})$ indicates the polarization of the Bloch wavefunctions in the Chern-sublattice basis, where $\tilde{\sigma}_z(\boldsymbol{k}) = +1 \,(-1)$ corresponds to full polarization in the A (B) basis. **b,** Same as (a) except in the limit where inter-Chern tunneling has been switched off. **c,** Stoner model analysis of the sublattice bands in (b). Color indicates the difference in total occupations of the A and B sublattice bands, as a function of interlayer potential $\Delta$ and total filling factor $\nu$. Interaction strength $U_{\text{int}} = 20$ meV and twist angle $\theta = 1.8°$.

A useful basis for describing the strongly-interacting physics of HTG is the Chern-sublattice basis (also referred to just as the Chern basis or the sublattice basis). This is obtained by diagonalizing the microscopic sublattice operator $\sigma_z$ projected to the central bands. This yields a basis of moiré Bloch wavefunctions $|\psi_{\tau s \sigma}(\boldsymbol{k})\rangle$ where $\tau = K, K'$ indexes the valleys, $s = \uparrow, \downarrow$ indexes the spins, and $\sigma = A, B$ indexes the sublattice bands within each flavor. Note that this sublattice basis is distinct from the microscopic sublattice degree of freedom. However, the sublattice basis labelled by $\sigma$ is associated with Bloch wavefunctions that have a large, but not complete, polarization on the microscopic sublattice $\sigma$. For any spin, the $KA$ band has Chern number $C = +1$, while the $KB$ band has Chern number $C = -2$. The Chern numbers for the other valley are related by time-reversal. In the strong-coupling theory of HTG at integer fillings, the interacting physics is described by generalized ferromagnetism within the sublattice bands.

In Fig. S6a, we color code the central bands of HTG based on their polarization $\tilde{\sigma}_z(\mathbf{k})$ in the sublattice basis, where red (blue) corresponds to a large overlap with the $A$ ($B$) basis. Figure S6b shows the band structure where the inter-Chern tunneling within the central bands is switched off. This is achieved by expressing the kinetic term, projected to the flat bands, in the sublattice basis, and setting the off-diagonal components to zero. Figure S6b illustrates that the A-band is very narrow and has a higher average energy than the B-band. However, the B-band is dispersive enough that it contains pockets in momentum space with higher energy than the A-band.

**Stoner model for sublattice transitions**

Prior to performing scHF calculations, we consider a simplified Stoner model analysis. To capture the sublattice transition physics, we work with the sublattice bands of the continuum model where the inter-Chern tunneling is switched off (Fig. S6b). This leads to eight generalized flavors of electrons indexed by the composite label $f = (\tau, s, \sigma)$, which collects the valley, spin, and sublattice indices. The displacement field is introduced with layer potentials $\Delta, 0, -\Delta$ for simplicity. In the Stoner model, each flavor is characterized by its partial filling $\nu_f \in [0,1]$. Within each flavor, filling the Bloch states in order of increasing kinetic energy leads to the total kinetic energy function $E_{kin,f}(\nu_f)$. We introduce an on-site density-density interaction, of strength $U_{int}$, that is completely flavor-independent. The total energy of the system is then

$$E_{tot}(\{\nu_f\}) = \sum_f E_{kin,f}(\nu_f) + U_{int} \sum_{f \neq f'} \nu_f \nu_{f'}.$$

Note that this analysis neglects the properties of the Bloch functions and only depends on the energy-dependent density-of-states of each flavor. For each total filling factor $\nu = \sum_f \nu_f - 4$, the partial fillings are obtained by minimizing $E_{tot}(\{\nu_f\})$.

Figure S6c shows a phase diagram of the sublattice polarization as a function of interlayer potential $\Delta$ and filling factor $\nu$. Similar to the scHF calculation, the Stoner model predicts a nearly periodic sequence of sublattice transitions at $\Delta = 0$. We describe the nature of the transitions for an interaction strength $U_{int} = 20$ meV. At an integer filling $\nu = \nu' \geq 0$, the system fully occupies all B-bands and $\nu'$ A-bands, and begins to populate one A-band as $\nu$ increases. At a critical filling $\nu' < \nu_c < \nu' + 1$, the system suddenly completely fills this A-band, and partially depopulates all four B-bands. Then, the system repopulates these B-bands until the next integer filling $\nu = \nu' + 1$, where the process repeats itself. Note that the partial depopulation of all four B-bands above $\nu_c$ is not observed in the scHF analysis, and would imply a first-order transition, which is inconsistent with the experiment. As the displacement potential $\Delta$ is increased, the sublattice transitions in the Stoner model shift towards smaller $\nu$, which is also seen in the scHF calculations.

**Self-consistent Hartree-Fock calculations and sublattice transitions**

To gain theoretical understanding of the phase diagram of HTG as a function of $\nu$ and $D$, we perform scHF calculations on the continuum model of moiré-periodic h-HTG at $\theta = 1.8°$. Given the large gap to the remote bands, we project our calculations to the central two moiré bands per spin and valley. We incorporate dual-gate screened interactions $V(q) = \frac{e^2}{2\epsilon_0 \epsilon_r q} \tanh q d_{sc}$ with gate screening length $d_{sc} = 25$ nm. The overall interaction scale is controlled by the relative permittivity $\epsilon_r$, which accounts for the hBN dielectric environment and screening from remote bands. Since the value of $\epsilon_r$ is not known precisely, we will treat it as a theoretical tuning parameter in the calculations. We use the 'average' interaction scheme, where the four-fermion term is normal-ordered relative to the average density of the central bands. We restrict to spin-collinear density matrices, but allow for intervalley coherence at spiral wavevectors $\boldsymbol{q}$ corresponding to any of the $\gamma, \kappa, \kappa'$ points. Note our interaction term is density-density in flavor space, such that the Hamiltonian obeys an enhanced $U(2)_K \times U(2)_{K'}$ flavor symmetry. Therefore, our scHF calculations are not able to fully resolve the energy competition between different flavor orderings. This enhanced symmetry will be reduced to the physical $SU(2)_s \times U(1)_v \times U(1)_c$ symmetry by various 'Hunds coupling' terms that are neglected in the present analysis.

To aid the interpretation of the scHF results, we evaluate the average occupation numbers $v_i^\sigma \in [0,1]$ of the scHF density matrix in the sublattice basis, where $i = (\tau, s)$ indexes the four flavors. Representative results are exhibited in Fig. S7 for different interaction strengths. We first describe an intuitive framework for describing the physics at $\Delta = 0$ in terms of the sublattice bands. The ground state at charge neutrality $\nu = 0$ consists of fully occupied B-band within each flavor. This can be understood from a combination of the tendency towards sublattice (exchange) ferromagnetism, and the fact that the B-bands have an overall lower kinetic energy than the A-bands. When electrons are initially doped into the system, they enter the A-bands. Because the A-bands are very narrow, the system polarizes the flavor of these additional carriers. However, as the filling factor is further increased, the system experiences a sequence of sublattice transitions that repeat with a period of roughly $\Delta \nu = 1$. For example, these occur at $\nu_c \approx 0.4, 1.4, 2.4, 3.4$, for $\epsilon_r = 11$. Below one such sublattice transition at $\nu_c$, the system is partially populating an A-band. Around $\nu = \nu_c$, the system rapidly completes the occupation of this A-band, at the expense of partially emptying one or more B-bands. Above $\nu_c$, the system repopulates these B-bands, until they get fully occupied at the next integer filling. Subsequently, the sublattice seesaw transitions repeat themselves in the rest of the flavors. For each non-negative integer filling, the scHF calculation predicts a correlated insulator, which further necessarily breaks time-reversal at $\nu = 1$ and 3.

Before analyzing the transitions at $\Delta = 0$ in more detail, we briefly discuss the physics for finite $\Delta$. As the interlayer potential is increased, the sublattice transitions drift to smaller densities (Fig. S7). For small values of $\Delta$, the nature of the sublattice transitions remains unchanged from $\Delta = 0$. For sufficiently large $\Delta$, the sublattice transitions can cross integer fillings. However, a robust physical interpretation of the sublattice transitions for large $\Delta$ is challenging. First, the scHF ground states for large $\Delta$ significantly hybridize the sublattice bands, which complicates any picture based on simply occupying sublattice bands. Furthermore, the flavor symmetry-breaking becomes more complex. For instance, the middle two columns of Fig. S7 deviate for $\Delta \geq 40$ meV due to the emergence of intervalley coherence. Additionally, the proposed topological transitions at $\nu = 1, 2$ in Ref. [15] occur at similar $D$ values to where the sublattice transitions cross integer fillings. At present, however, we do not have any physical argument that would require these two phenomena—topological phase transitions and seesaw transitions—to coincide at the same $D$. For this reason, we refrain from drawing a definitive conclusion in the present work.

We note that while the scHF calculations robustly reproduce the overall trajectories of the seesaw transitions, they do not capture all fine details of the experimental phase diagram. In particular, in the experimental data feature III exhibits a sharp bend toward lower filling as it crosses $\nu = 2$, which is less pronounced in the calculations and has no obvious counterpart for feature II near $\nu = 1$. Nevertheless, the scHF results reveal a qualitative distinction between these two transitions at large $\Delta$: above the transition corresponding to feature III, the Hartree–Fock ground state becomes intervalley coherent, whereas no such change in electronic order occurs above feature II. This difference may underlie the distinct curvature of feature III near $\nu = 2$. A complete microscopic understanding of this behavior, however, lies beyond the scope of the present modeling and requires further theoretical investigation.

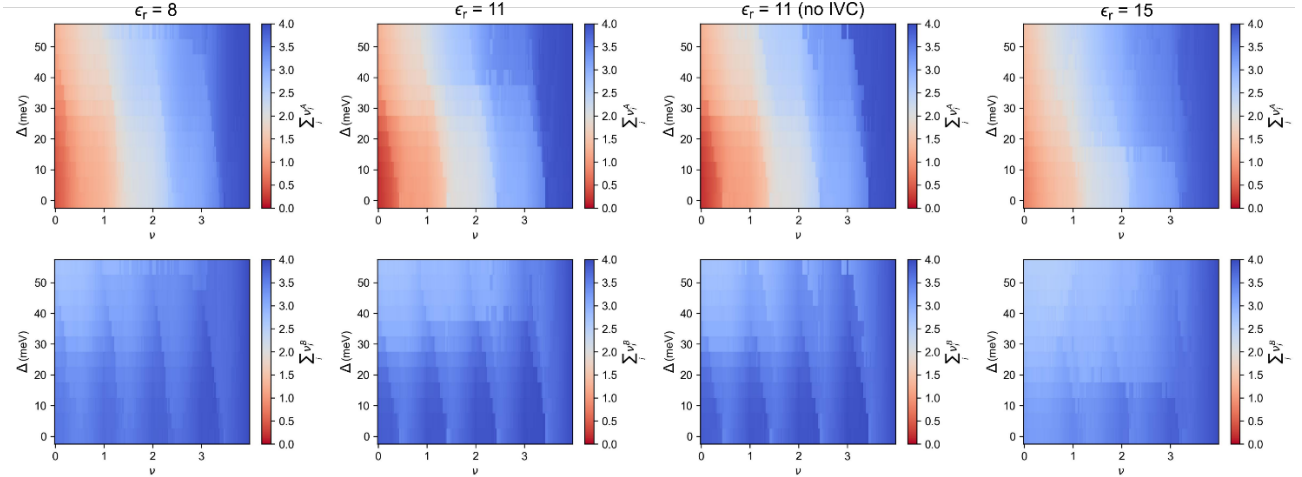

**Fig. S7. scHF phase diagrams for different Coulomb interaction strengths.** Top (bottom) row shows the total occupation of the A (B) sublattice basis summed over all flavors, as a function of the total filling factor $\nu$ and interlayer potential $\Delta$. Different columns correspond to varying interaction strengths, parameterized by $\epsilon_r$. For $\epsilon_r = 11$, we also repeat the calculations with intervalley coherence disallowed. System size is 12×12.

Additionally, we do not have any physical argument that these two features (topological phase transition and seesaw transition) should happen at the same *D*. For this reason, we do not draw a definitive conclusion in the present work.

We return to an explanation of the physics near $\Delta = 0$. There are two key ingredients that drive the sublattice transitions: sublattice exchange and the asymmetric dispersion of the sublattice bands. Inspired by the numerical scHF results, we illustrate the interplay of these aspects with a toy model that consists of a single pair of A- and B-bands (see Fig. 4A). For simplicity, we consider the limit where the inter-Chern tunneling is switched off, so there is no mixing between the two sublattice bands. Sublattice exchange, which is important for strong interactions, dictates that the system should try maximizing the sublattice polarization. Furthermore, the exchange energy of a filled A-band is nearly the same as that of a filled B-band as pointed out in the strong-coupling theory [18], so the competition between the sublattices is essentially determined by kinetic energy. At half-filling (i.e. $\nu = 1$) of this pair of bands, the system chooses to fully occupy the B-band since it has a lower overall kinetic energy. When electron-doping $\nu > 1$, the added carriers have no choice but to populate the empty A-band, thereby forming an A-metal (with a fully occupied B-band). However, hole-doping of $\nu = 2$ state, where both sublattices are fully filled, leads to a different conclusion. In particular, the B-band has a dispersive pocket that pokes above the entire A-band, so that holes prefer to enter the B-band and form a B-metal (with a fully occupied A-band). Therefore, there exists a critical filling $1 < \nu_c < 2$ where the system transitions from A-metal to B-metal. The critical filling can be estimated from when the relative total kinetic energies of A-metal and B-metal change signs, as illustrated in main text Fig. 4I. Figure S8 shows representative scHF band structures for $2 < \nu < 3$, highlighting the sublattice transition that occurs at $\nu_c \approx 2.4$ for $\epsilon_r = 11$ and $\Delta = 0$. Note that there is a finite amount of hybridization between the sublattice bands, and the band width of the partially filled metallic bands is enhanced in scHF. However, it is evident that the transition involves a redistribution of carriers between the sublattices, with corresponding exchange-induced shifts in the sublattice bands. Exact diagonalization studies in the filling range $3 < \nu < 4$ have also observed a sublattice transition [19], implying the robustness of this phenomenon beyond mean-field theory.

The shift of the sublattice transitions in Fig. S7 to smaller densities with increasing $\Delta$, at least for small $\Delta$, can be understood by considering the response of the A- and B-bands to an interlayer potential. At $\Delta = 0$, the in-plane $C_{2y}$ rotation symmetry forbids a net layer polarization of the sublattice bands. This means that the average kinetic energy of each sublattice band is relatively insensitive to $\Delta$, so that the intercepts of the curves in Fig. 4I at $\nu = 1$ are only weakly affected. Despite this, the B-bands carry a significant momentum-contrasting layer dipole moment, unlike the A-bands [18]. Therefore, a finite interlayer potential substantially increases the dispersion of the B-bands and reduces $\nu_c$.

Note also that the seesaw transitions occur only upon electron doping due to the kinetic energy competition between A- and B-bands, consistent with the experiment. Upon hole doping, such a competition does not exist because the partially filled valence B-bands always have lower kinetic energy.

Finally, we address the flavor physics at the sublattice transitions. For $\epsilon_r = 11$, the seesaw switching happens within a single flavor as shown in Fig. S8. This is compatible with a continuous transition as there is no discontinuous jump in the flavor polarization. The filling interval over which a sublattice switch occurs is narrower for stronger interactions. For sufficiently strong interactions though, the sublattice transition qualitatively changes and becomes associated with a simultaneous flavor jump. This happens because exchange ferromagnetism prefers to maximally polarize as many flavors as possible, even if it is kinetically unfavorable. For example, for the $\nu_c \approx 2.4$ transition at $\epsilon_r = 8$, the B-band that gets partially emptied belongs to a different flavor (in particular the one that was half-filled just below $\nu_c$). Such a transition would necessarily be first-order, and hence incompatible with the continuous magnetization transitions observed experimentally (Fig. S5). A different complication occurs for sufficiently weak interactions. In this regime, interactions are not strong enough to flavor-ferromagnetize the dispersive pockets of the B-bands, so that above $\nu_c$, metallic Fermi surfaces in the B-band are generated across multiple flavors. This would again be a first-order transition, and inconsistent with the experiment. Hence, we believe that the intermediate-coupling regime ($\epsilon_r \approx 11$) in the scHF calculations is the most consistent with experimental phenomenology. The full scHF results for the band structure as a function of filling for $\epsilon_r = 11$ and $\Delta = 0$ is shown in Movie S1.

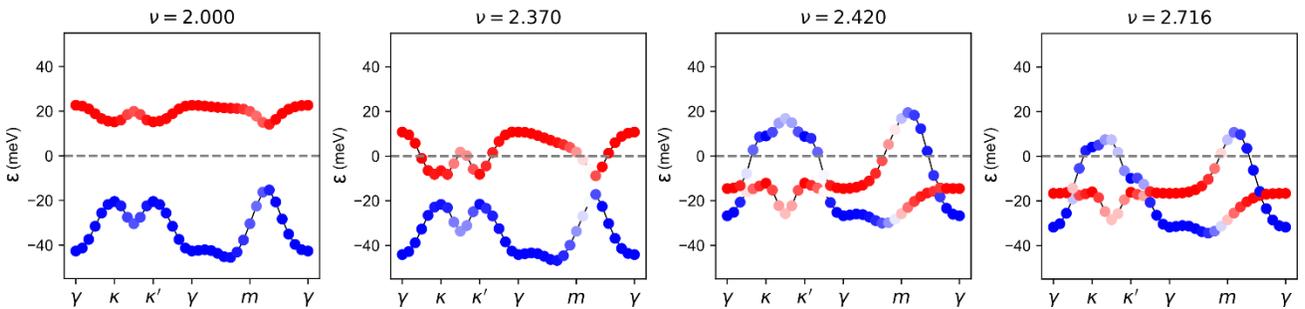

**Fig. S8. Mechanism for sublattice transitions.** Self-consistent HF band structure at interaction strength $\epsilon_r = 11$ and interlayer potential $\Delta = 0$ for fillings below and above the critical $\nu_c \approx 2.4$. The bands are color-coded according to their polarization in the sublattice basis, with red (blue) indicating the A (B) basis. Only the bands of the active flavor with Fermi surfaces are shown. System size is 18×18.

**Toy model for orbital magnetization of sublattice bands**

In this section, we introduce a toy model for studying orbital magnetization in interacting HTG, focusing on the contributions arising from the sublattice physics. As recently shown, the single particle equations for the orbital magnetization can also be applicable to interacting scHF wave functions [54]. We first review the standard formulas for orbital magnetization in Bloch bands. Consider a Bloch Hamiltonian $H(\boldsymbol{k})$ that yields a set of bands $n$ with band energy $\varepsilon_n(\boldsymbol{k})$ and Bloch functions $|n\rangle$ (the Bloch momentum $\boldsymbol{k}$ is left implicit). We define the following quantities for convenience

$$U_{nm}(\boldsymbol{k}) = \frac{e}{\hbar} Im \frac{\langle n|\partial_{k_x} H(\boldsymbol{k})|m\rangle\langle m|\partial_{k_y} H(\boldsymbol{k})|n\rangle}{[\varepsilon_n(\boldsymbol{k}) - \varepsilon_m(\boldsymbol{k})]^2}$$

$$V_{nm}(\boldsymbol{k}) = \frac{e}{\hbar} Im \frac{\langle n|\partial_{k_x} H(\boldsymbol{k})|m\rangle\langle m|\partial_{k_y} H(\boldsymbol{k})|n\rangle}{[\varepsilon_n(\boldsymbol{k}) - \varepsilon_m(\boldsymbol{k})]},$$

where $U_{mn}(\boldsymbol{k})$ has dimensions of magnetic moment over energy, and $V_{mn}(\boldsymbol{k})$ has dimensions of magnetic moment. Note that the Berry curvature $\Omega_n(\boldsymbol{k})$ can be related to $U_{mn}(\boldsymbol{k})$ as

$$\Omega_n(\boldsymbol{k}) = 2Im\langle\partial_{k_x} n|\partial_{k_y} n\rangle = 2Im \sum_{m \neq n} \frac{\langle n|\partial_{k_x} H(k)|m\rangle\langle m|\partial_{k_y} H(\boldsymbol{k})|n\rangle}{[\varepsilon_n(\boldsymbol{k}) - \varepsilon_m(\boldsymbol{k})]^2} = 2\frac{\hbar}{e} \sum_{m \neq n} U_{nm}(\boldsymbol{k}).$$

The formulas for the self-rotation and the Chern magnetic moments of band $n$ are

$$m_n^{SR}(\boldsymbol{k}) = \frac{e}{\hbar} Im \langle\partial_{k_x} n|[H(\boldsymbol{k}) - \varepsilon_n(\boldsymbol{k})]|\partial_{k_y} n\rangle = -\sum_{m \neq n} V_{nm}(\boldsymbol{k})$$

$$m_n^C(\boldsymbol{k}) = -2\frac{e}{\hbar} Im \langle\partial_{k_x} n|[\varepsilon_F - \varepsilon_n(\boldsymbol{k})]|\partial_{k_y} n\rangle = -2[\varepsilon_F - \varepsilon_n(\boldsymbol{k})] \sum_{m \neq n} U_{nm}(\boldsymbol{k})$$

where $\epsilon_F$ is the Fermi level and in the last equalities, we have used the identity

$$|\partial_{k_i} n\rangle = \sum_{m \neq n} \frac{\langle m|\partial_{k_i} H(\boldsymbol{k})|n\rangle}{\varepsilon_n(\boldsymbol{k}) - \varepsilon_m(\boldsymbol{k})} |m\rangle.$$

We now consider a deformed Hamiltonian $\widetilde{H}(\boldsymbol{k})$ with momentum-independent but band-dependent shifts $\delta_n$

$$\widetilde{H}(\boldsymbol{k}) = H(\boldsymbol{k}) + \sum_n \delta_n |n\rangle\langle n|$$

$$\tilde{\varepsilon}_n(\boldsymbol{k}) = \varepsilon_n(\boldsymbol{k}) + \delta_n.$$

Importantly, the shifts are introduced in the same basis that diagonalized the original Hamiltonian $H(\boldsymbol{k})$, so that the Bloch functions $|\tilde{n}\rangle = |n\rangle$ remain unchanged. The new self-rotation and Chern magnetic moments are

$$\widetilde{m}_n^{SR}(\boldsymbol{k}) = m_n^{SR}(\boldsymbol{k}) - \delta_n \frac{e}{2\hbar} \Omega_n(\boldsymbol{k}) + \sum_{m \neq n} \delta_m U_{nm}(\boldsymbol{k})$$

$$\widetilde{m}_n^C(\boldsymbol{k}) = m_n^C(\boldsymbol{k}) + \delta_n \frac{e}{\hbar} \Omega_n(\boldsymbol{k}).$$

We now apply the above formalism to a toy model treatment of interacting HTG. We divide the bands of HTG into remote valence (r.v.), remote conduction (r.c.), and A and B sublattice bands (indexed by $\sigma$). Note that the spin-valley flavors indices are left implicit. In the limit where the inter-Chern tunneling is switched off, these bands diagonalize the non-interacting Hamiltonian. In the filling range of interest, the remote valence (conduction) bands are always fully occupied (unoccupied). To capture the sublattice exchange physics obtained in the scHF calculations, we allow for arbitrary energy shifts $\delta_\sigma$ of the sublattice bands, which depend on the spin-valley flavor. The corresponding total orbital magnetization is

$$M(\mu, \{\delta\}) = M^{SR}_{r.v.}(\{\delta\}) + M^{C}_{r.v.}(\varepsilon_F) + M^{SR}_{sub.}(\varepsilon_F, \{\delta\}) + M^{C}_{sub.}(\varepsilon_F, \{\delta\})$$

$$M^{SR}_{r.v.}(\{\delta\}) = -\sum_{k, r \in r.v.} \sum_{m \neq r} V_{rm}(\boldsymbol{k}) + \sum_\sigma \delta_\sigma \sum_{k, r \in r.v.} U_{r\sigma}(\boldsymbol{k})$$

$$M^{C}_{r.v.}(\varepsilon_F) = 2 \sum_{k, r \in r.v.} [\varepsilon_n(\boldsymbol{k}) - \varepsilon_F] \sum_{m \neq r} U_{rm}(\boldsymbol{k})$$

$$M^{SR}_{sub.}(\varepsilon_F, \delta) = \sum_{k, \sigma} f_{\sigma,\mu}(\boldsymbol{k}) \left[ -\sum_{m \neq \sigma} V_{\sigma m}(\boldsymbol{k}) - \delta_\sigma \sum_{m \neq \sigma} U_{\sigma m}(\boldsymbol{k}) + \delta_{\bar\sigma} U_{\sigma\bar\sigma}(\boldsymbol{k}) \right]$$

$$M^{C}_{sub.}(\varepsilon_F, \delta) = \sum_{k, \sigma} f_{\sigma,\mu}(\boldsymbol{k}) \left[ 2[\varepsilon_\sigma(\boldsymbol{k}) + \delta_\sigma - \varepsilon_F] \sum_{m \neq \sigma} U_{\sigma m}(\boldsymbol{k}) \right]$$

where $f_{\sigma,\mu}(\boldsymbol{k})$ is the zero-temperature occupation factor for sublattice band $\sigma$, which depends on $\varepsilon_F, \varepsilon_\sigma(\boldsymbol{k}), \delta_\sigma$ (all orbitals with shifted energies below the Fermi level $\varepsilon_F$ are occupied). The orbital magnetization can therefore be computed by specifying nine energies, which consist of $\varepsilon_F$, and the eight sublattice shifts $\delta_\sigma$ across the four spin-valley flavors. Equivalently, as discussed in the main text, one can think of the eight sublattice shifts $\delta_\sigma$ as eight individual chemical potentials $\mu_i$ where $i$ indexes the eight generalized flavors $i = (\tau, s, \sigma)$, along with the overall Fermi level $\varepsilon_F$.

For a given density, $\nu$, we choose the nine energies based on insights taken from the scHF study (Movie S1). As discussed, the intermediate coupling regime ($\epsilon_r = 11$) best matches the experimental phenomenology, so we choose that only a single flavor is at the Fermi level between consecutive integer fillings $\nu'$ to $\nu' + 1$ ($\nu' = 0,1,2,3$), and equivalently that the compressible sublattice transition swaps between an A- and B-bands of the same flavor. Empty (full) bands are shifted by $+20$ ($-20$) meV consistent with scHF. Next, similar to scHF, $\varepsilon_F(\nu)$ is chosen to be sawtooth shaped, where at integer filling the Fermi level jumps up to the next A-band, and then gradually lowers into the gap as the band is doped. The overall amplitude of $\varepsilon_F(\nu)$ is taken to be smaller than in scHF, since if $\varepsilon_F$ varies too much, the magnetic signal from the Chern insulating states at $\nu = 1, 3$ quickly become much larger than the magnetic signal arising from the sublattice transition, which is in stark contrast to experiment. We therefore posit that $\epsilon_F$ doesn't change that much and is relatively constant at the sublattice transitions (which is also consistent with a second order transition) as discussed in the main text. This might be explained by spatial disorder in the system which causes finite states to appear in the gap, relatively pinning $\varepsilon_F$ to a limited portion of the gap. The sequence of flavors to be filled is chosen to be $K \uparrow$, $K' \downarrow, K' \uparrow, K \uparrow$ such that the system has a $C = 0$ gap at $\nu = 2$, reflecting the absence of AHE at $\nu = 2$. Finally, the critical transition densities $\nu_c$ are chosen to be $0.5, 1.3, 2.4, 3.4$ following experiment. The resulting band structure as a function of $\nu$ is shown in Movie S2. $M(\nu)$ is calculated using the above formalism and shown in Fig. 4D and Movie S3.

We comment that our toy model incorporates interactions by simply shifting the overall energy of the sublattice bands in the non-interacting Hamiltonian. It neglects the mixing between the sublattice bands, and the momentum-dependent renormalization of the bands due to interaction effects. Indeed, this simplified model cannot fully account for all the subtleties of an interacting system and is employed merely to gain an intuitive understanding of what causes the large jumps in magnetization upon the seesaw transitions. A more microscopic treatment would require a general theoretical framework for addressing orbital magnetization in interacting systems, which is beyond the scope of the present study.

**Null hypothesis**

A natural starting point for interpreting the magnetization data is the standard assumption that interactions act only within the four conduction (A) bands, while the four valence (B) bands remain inert. This picture, familiar from many moiré systems, provides a baseline or null hypothesis against which more complex scenarios can be judged. In this section we examine this null hypothesis in detail and show that it fails both qualitatively and quantitatively to reproduce the experimental observations.

To enforce this scenario microscopically, we artificially freeze out the B bands by adding to them a large negative energy shift $\Delta E_B$. In this limit, only the A bands remain active, and scHF calculations predict a simple "strong-coupling" behavior in which the four A bands fill sequentially and remain 1-fold degenerate (Fig. S9c). We verified that this behavior is robust over a wide range of parameters, including very small interaction strength ($\varepsilon_r = 24$) and finite displacement field ($D = 20$ meV). The origin of this trivial filling sequence lies in the extreme flatness of the A bands, which suppresses kinetic-energy competition and allows exchange physics to dominate, leading to a simple 1-fold sequence. Figure S9d shows the calculated total magnetization, $M_z$, for this scenario (all magnetization calculations follow the same procedure outlined in the main text and Supplementary text). As expected, large magnetization jumps are present only at integer fillings, where the chemical potential crosses a Chern gap and the Chern magnetization $M_C$ changes abruptly. No magnetization jumps appear in the compressible states, because no electronic reordering occurs there. This is seen clearly in the differential magnetization in Fig. S9e. Importantly, the magnetization is almost constant in the interval $0 < \nu < 1$ as $M_{SR}$ of the A bands is quite small, less than 0.1 μ_B/u.c. (Fig. S9a) as compared to $M_C$ contribution of more than 2 μ_B/u.c. upon crossing a Chern gap (Fig. S9d).

To more robustly check if our data might possibly be explained by the null hypothesis, we assume that a 1-fold A sequence persists similar to the scHF calculation, but we additionally enforce time-reversal transitions at densities observed in experiment (i.e. $\nu = 0.5, 1.3, 2.4, 3.4$). Such sequence of transitions is shown in Fig. S9f. The calculated total and differential magnetizations are presented in Figs. S9g,h respectively. In this scenario magnetic jumps occur in the compressible state with peaks in the differential magnetization (Fig. S9h), however two clear discrepancies remain compared to the experiment (Fig. S9b). First, the computed magnitudes of the incompressible (gaps) and compressible peaks are similar, whereas experimentally the compressible peaks are larger than the incompressible ones. Second, only two compressible peaks (II and IV) appear instead of the four observed in experiment.

Of course, transitions between A bands need not be limited to time-reversal exchanges between the $K$ and $K'$ valleys. Other systems exhibit degeneracy transitions driven by their specific fermiology. For example, rhombohedral graphene exhibits 1-fold to 2-fold and 2-fold to 4-fold transitions as a mechanism to avoid a van-Hove singularity [36], and cascade transitions have been proposed in MATBG due to the zero-energy Dirac-cone structure of its bands [33,34]. Although our scHF calculations do not generate such fermiology-driven transitions, we nevertheless test scenarios incorporating them to assess whether they could reproduce the experimental data.

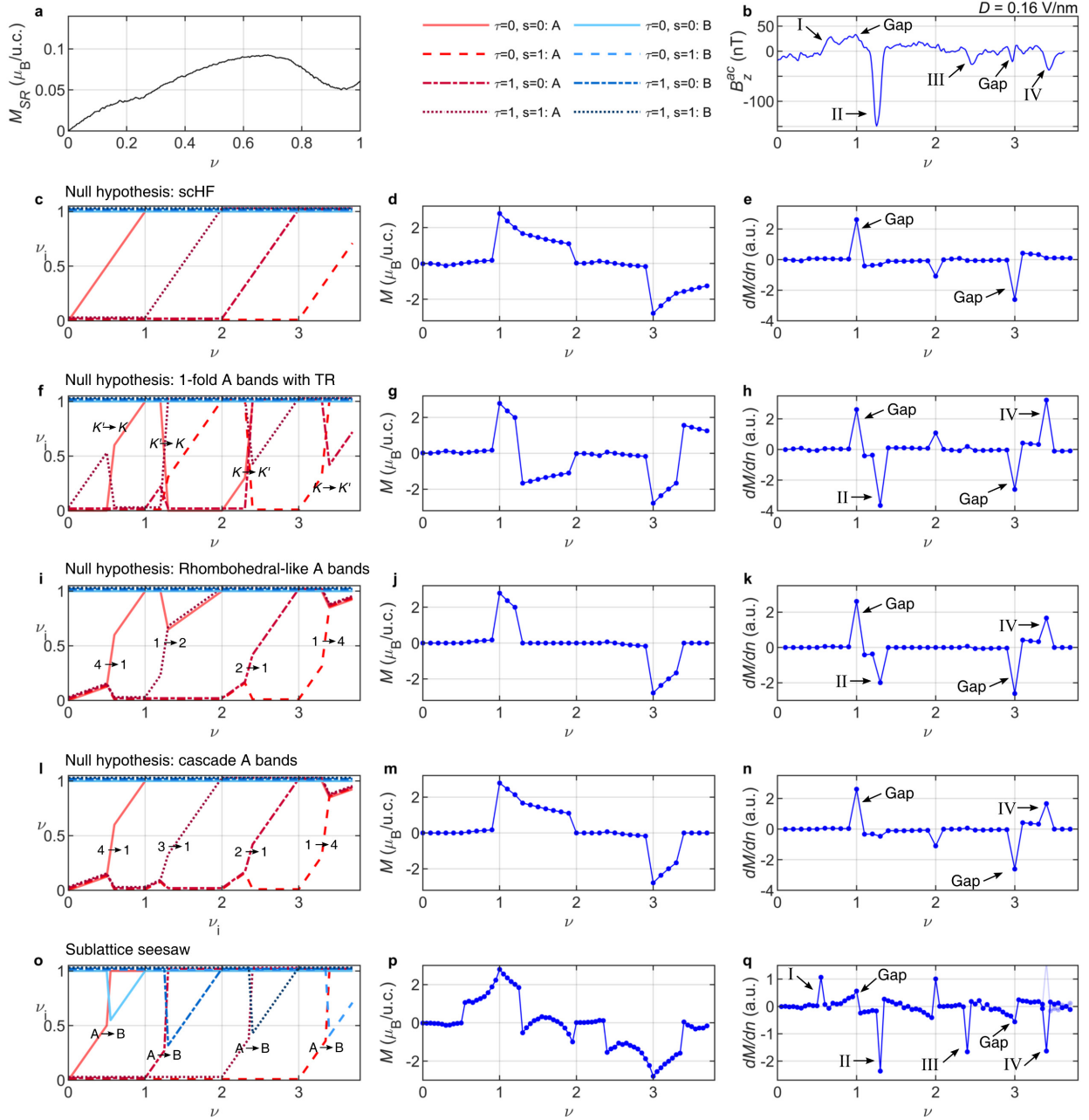

**Fig. S9. Analysis of the null hypothesis. a,** Self rotation magnetization, $M_{SR}$, as a function of filling an A band. **b,** Linecut of experimental $B_z^{ac}$, reproduced from Fig. 4F. **c-e**, scHF results while artificially freezing out B bands by adding a large negative energy shift $\Delta E_B$. Flavor resolved filling as a function of $\nu$ is shown in (**c**), along with calculated total magnetization (**d**) and differential magnetization (**e**). **f-h**, Same as (c-e) but with additional time reversal transitions between $K$ and $K'$ bands enforced artificially at the transition densities observed in experiment (i.e. $\nu = 0.5, 1.3, 2.4, 3.4$). The type of transition is labeled schematically in (**f**). **i-k**, Same as (f-h) but for a different null hypothesis. Here transitions are between states of different degeneracies (labelled schematically in (**i**), similar to transitions seen in rhombohedral stacked graphene. **l-n**, Same as (f-h) but for a yet another null hypothesis. Here transitions are similar to cascade transitions previously proposed in MATBG. **o-q**, Same as (f-h) but for the sublattice seesaw transitions proposed in the main text. This scenario is the only one that can reproduce compressible magnetic peaks larger than incompressible peaks, in addition to the four compressible peaks observed in experiment.

Figure S9i shows a scenario modeled after rhombohedral graphene, in which the system transitions between 1-fold, 2-fold, and 4-fold degenerate states. As before, the transitions are enforced at the densities observed experimentally. The sequence is chosen to maintain time-reversal symmetry at $\nu = 2$, consistent with the data, and we assume that the system is 4-fold degenerate near $\nu =0$ and 4, where interactions are generally weaker. Figures S9j,k display the resulting total and differential magnetizations. Once again, the same two discrepancies persist relative to experiment: the compressible and incompressible peaks have comparable magnitudes, and only two compressible peaks are produced.

Finally, we test a Dirac-revival–type scenario (Figs. S9l-n). In this construction, all A bands are either completely full or empty at each integer filling. The empty bands remain degenerate upon low doping, and are then forced into a 1-fold state at the experimentally observed compressible-transition densities (Fig. S9l). The resulting total and differential magnetizations (Figs. S9m,n) again fail to reproduce the experimental behavior: the compressible and incompressible peaks have similar magnitudes, and no more than two compressible jumps appear.

In fact, it is not a coincidence that, even when four compressible transitions are explicitly enforced, none of the tested null hypotheses could reproduce all four compressible magnetization jumps. The underlying reason is a fundamental limitation of any A-only model. As described in the main text and shown in Figs. 4B,C, interacting bands that shift in energy relative to one another, can be described by their individual chemical potentials $\mu_i$. Since $M_{SR}$ of the A bands is negligible (Fig. S9a), a large magnetization jump requires a large change in the $\mu_i$ of an A band leading to a large $M_C$ contribution. Such a large change in $\mu_i$ can occur only if a band is rapidly filled and pushed far below the Fermi level (positive jump in $\mu_i$) or if a band below the Fermi level is rapidly emptied (negative jump in $\mu_i$), as illustrated in Fig. 4B for the seesaw mechanism. For $\nu < 1$, no A band can be fully occupied and driven below the Fermi level; consequently, any rearrangement among partially filled A bands produces only small shifts in $\mu_i$, failing to generate a large magnetization peak at transition I. Thus, any A-only scenario does not reproduce transition I and cannot reproduce the four observed experimental peaks.

On the other hand, the sublattice seesaw, which allows non-trivial interplay between the A and B bands, reproduces all the experimental features (Fig. S9o-q): the four transitions appear naturally due to alternation between A and B bands for each flavor, the largest jumps occur in the compressible states, and despite the negligible $M_{SR}$ of the A bands, all four transitions produce significant magnetization jumps.

**Minor hysteresis loop measurements**

Figure S10 shows $B_z^{ac}$ measurements in device B acquired upon sweeping $\nu$ within limited range between $\nu_{min}$ and $\nu_{max}$ in the vicinity of $\nu = 3$. We first set $\nu_{min}$ to 2.1 below seesaw transition line III and sweep $\nu$ up to $\nu_{max} = 3.2$ and back to $\nu_{min}$. Figure S10a (bottom curve) shows no hysteresis upon this sweep. We then increment $\nu_{max}$ in small steps. The hysteresis sets in abruptly only when $\nu_{max}$ reaches or exceeds the seesaw transition line IV (top two curves). In particular, the differential magnetization peak at $\nu = 3$ is negative and shows no hysteresis as long as $\nu_{max}$ remains below line IV and turns hysteretic once $\nu_{max}$ exceeds line IV. We then repeat these measurements by fixing $\nu_{max}$ at 3.7 and sweeping $\nu$ down to $\nu_{min}$ and back. No hysteresis is observed as long as $\nu_{min}$ is above seesaw transition line III (bottom two curves in Fig. S10b), showing a positive differential magnetization peak at $\nu = 3$. Once $\nu_{min}$ crosses seesaw line III the hysteresis reappears (top two curves).

These measurements indicate that the hysteretic TR transitions between $K$ and $K'$ valleys are triggered by seesaw transitions, rather than by crossing the Chern insulator state at $\nu = 3$, as is the common case. Moreover, since $K$ and $K'$ valleys have opposite Chern numbers, two seesaw transition lines are expected to display positive differential magnetization peaks and two negative peaks. Figure 4E indeed shows positive peaks at transitions I and IV (light blue) and negative peaks at transitions II and III. In contrast, in most of the measurements, like in Fig. 2B, peak I is positive (blue) while the rest three peaks are negative (red). This apparent discrepancy is caused by the hysteretic TR transitions. The top blue curve in Fig. S10b shows that on approaching seesaw line IV, a positive differential magnetization peak, marked by the blue arrow, starts to develop, but the triggering of TR transition abruptly flips the sign of the magnetization signal. Such positive precursor peak is absent in the two bottom curves in absence of hysteresis. The same qualitative behavior is observed upon approaching seesaw line III from above, as indicated by the red arrow in Fig. S10a. This suggests that while the seesaw transitions reflect second-order sublattice interchange within a single flavor, they can also trigger a first-order hysteretic switching between flavors with opposite valley, as illustrated schematically in Fig. 3E. The blue line in Fig. 4E shows the calculated differential magnetization taking into account the TR transition at seesaw line IV. More theoretical and experimental work is needed to understand the mechanisms relating the sublattice seesaw transition and hysteretic switching.

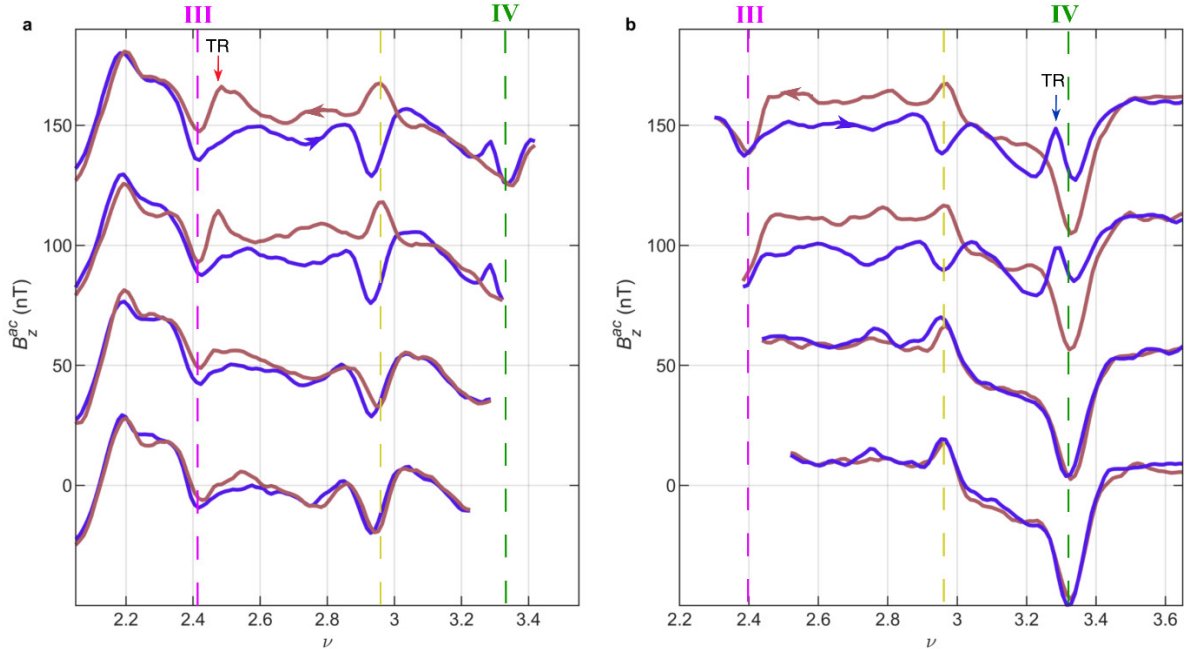

**Fig. S10. Local minor hysteresis loops. a,** Minor local hysteresis loops of $B_z^{ac}$ in device B upon sweeping $\nu$ from 2.1 to $\nu_{max}$ (blue) and back (red) while incrementing $\nu_{max}$ from 3.2 to 3.4 at $D = 0.177$ V/nm. The consecutive curves are offset by 50 nT for clarity. Magenta and green dashed lines indicate seesaw transition lines III and IV respectively, and yellow dashed line indicates the Chern insulating state. $B_z^{ac}$ becomes hysteretic once $\nu_{max}$ reaches peak IV. The red arrow marks TR transition upon sweeping $\nu$ down. **b,** Same as (a) but initializing the system at high $\nu$ and decrementing $\nu_{min}$ from 2.5 to 2.3. $B_z^{ac}$ becomes hysteretic once $\nu_{min}$ reaches peak III. The blue arrow marks the TR transition upon sweeping $\nu$ up.

**Movie Captions**

**Movie S1. Evolution of the scHF band structure with doping.** Band structure derived from scHF calculations at interlayer potential $\Delta = 0$ and interaction strength $\epsilon_r = 11$, which best matches the experimental phenomenology (Supplementary Materials). The dashed line marks the Fermi level. Since the scHF calculations have an enhanced $U(2)_K \times U(2)_{K'}$ flavor symmetry, they do not differentiate between the different flavor orderings and hence the four sets of low-energy bands are not labeled. The bands are color-coded according to their polarization in the sublattice basis, with magenta (cyan) indicating the A (B) basis. Even though the A- and B-bands hybridize after the sublattice transition, the empty states are mostly polarized to the B sublattice basis following the transition (cyan), which motivates the use of the simplified sublattice polarized bands to model the orbital magnetization of the system (Movies S2 and S3).

**Movie S2. Model of band structure evolution with sublattice polarized bands.** Simplified band structure model as a function of doping inspired by scHF. Interactions rigidly shift the sublattice polarized bands, where tunneling between the A- and B-bands has been suppressed. For each density the shifts are chosen to follow experimental phenomenology (Supplementary Materials). The sublattice transitions in the compressible states periodically swap between an A- and B-band for each flavor. The corresponding $M(\nu)$ evolution is shown in Fig. 4D and Movie S3.

**Movie S3. Magnetization jumps at the sublattice transitions.** Calculated magnetization as a function of $\nu$ for a pair of A (cyan) and B (magenta) sublattice-polarized bands in the $K'$ valley, across a sublattice seesaw transition. The individual chemical potentials $\mu_{A,B}$ measured from the bottom of each band are shown, along with the self rotation ($M_{SR}$) and Chern ($M_C$) contributions to the total orbital magnetization ($M_{tot}$). Upon initial doping, the A-band and the Fermi level, $\varepsilon_F$, shift down, while the B-band remains fully occupied. The $M_{SR}$ of a full band is constant and independent of $\varepsilon_F$. In contrast, $M_C$ is determined by $\mu_i$ of each of the bands. Upon initial doping of the A-band, $\mu_A$ increases but $\mu_B$ decreases because $\varepsilon_F$ shifts down. At the seesaw transition, the A- and B-bands flip and $\mu_B$ drops sharply while $\mu_A$ jumps up. Since the bands have Chern numbers with opposite signs, the corresponding changes in $M_C$ of the two bands add up and create a large jump in $M_{tot}$.